\documentclass[aps,prx,showpacs,twocolumn,amsmath,amssymb,superscriptaddress]{revtex4}

\usepackage[colorlinks,citecolor=blue,linkcolor=cyan]{hyperref}
\usepackage[usenames,dvipsnames,svgnames]{xcolor}
\usepackage{Jonasmacros}
\usepackage{graphicx}
\usepackage{dcolumn}
\usepackage{bm}
\usepackage{mathrsfs}
\usepackage{subfigure}
\usepackage{natbib}
\usepackage{mathptmx}
\usepackage{pxfonts}
\usepackage{notoccite}
\usepackage{soul}

\begin{document}
\title{Time-dependent spin and transport properties of a single-molecule magnet in a tunnel junction}

\author{H. Hammar}
\affiliation{Department of Physics and Astronomy, Uppsala University, Box 530, SE-751 21 Uppsala}


\author{J. Fransson}
\email{Jonas.Fransson@physics.uu.se}
\affiliation{Department of Physics and Astronomy, Uppsala University, Box 530, SE-751 21 Uppsala}

\date{\today}

\begin{abstract}
In single-molecule magnets, the exchange between a localized spin moment and the electronic background provides a suitable laboratory for studies of dynamical aspects of both local spin and transport properties. Here we address the time-evolution of a localized spin moment coupled to an electronic level in a molecular quantum dot embedded in a tunnel junction between metallic leads. The interactions between the localized spin moment and the electronic level generate an effective interaction between the spin moment at different instances in time. Therefore, we show that, despite being a single spin system, there are effective contributions of isotropic Heisenberg, and anisotropic Ising and Dzyaloshinski-Moriya character acting on the spin moment. The interactions can be controlled by gate voltage, voltage bias, the spin polarization in the leads, in addition to external magnetic fields. Signatures of the spin dynamics are found in the transport properties of the tunneling system, and we demonstrate that measurements of the spin current may be used for read-out of the local spin moment orientation.
\end{abstract}


\maketitle
\section{Introduction}
\label{sec-intro}
Single-molecule magnets provide interesting workbench opportunities to study quantum phenomena related to their individual properties as well as promising potential for quantum information technology and quantum computation based on spintronics devices. Easy control of single magnetic moments paved the way for a deeper exploration of, e.g., magnetic anisotropies and exchange interaction, as well as new routes for significantly less energy consuming active electronics devices and information storage.

Molecular magnets offer a platform for studies of magnetic properties on a fundamental level due to their intrinsic discreteness. Experimentally, this has paved the way for electronic control and detection of the magnetization of individual molecules \cite{Hauptmann:2008aa,Loth:2010aa,Wagner:2013aa}, magnetic anisotropy, and exchange interaction of single atoms such as, e.g., Co and Mn on a surface \cite{Hirjibehedin19052006,PhysRevLett.98.056601,Meier04042008,PhysRevLett.102.257203,PhysRevB.78.155403}, and tuning of the magnetic anisotropy in molecular magnets \cite{PhysRevLett.114.247203}. Furthermore, spatial anisotropies have been observed for the Ruderman-Kittel-Kasuya-Yosida (RKKY) interaction \cite{Zhou:2010aa} as well as signatures of superexchange interaction and the long-range Kondo effect between single magnetic molecules \cite{PhysRevLett.101.197208,PhysRevLett.103.107203,Pruser:2011aa}. These advances in experimental techniques have led to realizations of magnetically stable atomic scale configurations \cite{Khajetoorians27052011,Loth13012012,Khajetoorians04012013} that are important steps toward the creation of stable magnetic memory devices at the atomic scale. Magnetic molecules containing transition metal atoms, e.g., M-phthalocyanine and M-porphyrins where M denotes a transition metal element (Cr, Mn, Fe, Co, Ni, Cu) \cite{Wende:2007aa,PhysRevLett.101.217203,PhysRevLett.110.157204,Raman:2013aa,Fahrendorf:2013aa}, as well as single molecules comprising complexes of transition metal elements \cite{PhysRevB.70.214403,ANIE:ANIE200352989} and antiferromagnetic rings \cite{CHEM:CHEM277,PhysRevB.67.094405,PhysRevLett.94.207208,PhysRevLett.98.167401,PhysRevLett.108.107204,PhysRevLett.104.037203} have been explored in many different contexts.

For technological applications, on the other hand, the potential of molecular magnets and magnetic materials is unlimited. A range of different spintronics devices have been proposed, both using spin currents \cite{Chappert:2007aa} or spin torque \cite{Locatelli:2014aa}. Such devices include molecular spin-transistors, molecular spin-valves, molecular multidot devices \cite{Bogani:2008aa}, etc. These can potentially be used both as building blocks of quantum computers \cite{Leuenberger:2001aa} and as quantum simulators \cite{PhysRevLett.75.729}. There are already several experimental realizations of these kinds of devices, including magnetic memories and spin qubits \cite{Mannini:2009aa,Timco:2009aa,Mannini:2010aa,PhysRevLett.97.207201}.

On the theoretical side, we have witnessed great progress over the course of the past decade in developments of the theory for, e.g., single molecular magnets and magnetization dynamics. There have been several studies of magnetic exchange interaction and the possibilities for electrical control of the interaction and spin transport \cite{PhysRevLett.108.057204,PhysRevLett.102.086601,PhysRevLett.111.127204,PhysRevB.77.205316,PhysRevB.87.045426,PhysRevLett.113.257201}. Under non-equilibrium conditions, magnetic molecules show signatures of intrinsic anisotropic exchange interactions that can be used to control molecular spin \cite{Misiorny:2013aa,PhysRevLett.111.127204}, something that may lead to read-and-write capabilities with currents in spintronics devices \cite{PhysRevB.73.235304,PhysRevB.75.134425,1367-2630-17-8-083020}. Non-equilibrium studies of transport properties have, moreover, suggested that vibrations coupled to the spin degrees of freedom may induce electrical currents that can provide interesting properties for, e.g., mechanical control of single magnetic molecules \cite{PhysRevLett.107.046802,PhysRevB.87.195136}. Superconducting spintronics also paves the way toward enhancing central effects of spintronics devices \cite{Linder:2015aa,PhysRevB.88.104512,PhysRevB.83.104521,PhysRevB.90.014516}.

The majority of the reported theoretical progress is, however, has been limited to stationary, or Markovian, processes. Although this is an important regime, both for fundamental studies as well as for technological applications, it is nonetheless crucial to control also transient properties induced by sudden on-sets and variation of the external conditions applied to the system. Regarding spin dynamics, the Landau-Lifshitz-Gilbert (LLG) equation is often postulated as the platform for theoretical studies, despite the fact that the (exchange and damping) parameters for this equation are typically taken on phenomenological grounds or from experiments. These parameters are, in addition, assumed to have a negligible time-dependence, something that cannot be taken for granted in nanoscale systems. Previous derivations of the LLG equation \cite{franssonNJP2008,PhysRevLett.108.057204} clearly illustrate that the electronically mediated exchange interactions depend strongly on the magnetization dynamics and are, hence, intrinsically dynamical quantities as well. The non-linearity of the dynamical equations indicates, moreover, that it is non-trivial to decide whenever the time-dependence of the interaction parameters can be neglected.

To begin to depart from the \emph{ad hoc} treatments of the dynamics of spins coupled to electron currents, in this paper we perform time-dependent studies and analyses beyond the Markovian and adiabatic approximations for both the spin-dynamics and the tunneling current. In addition, we include the interdependence between the current through the molecule and the localized magnetic moment by considering both action and back-action in the description. This can be regarded as the first loop in a self-consistent calculation, however, we do not perform our calculations to full self-consistency.

\begin{figure}[t]
	\centering
	\includegraphics[width=\columnwidth]{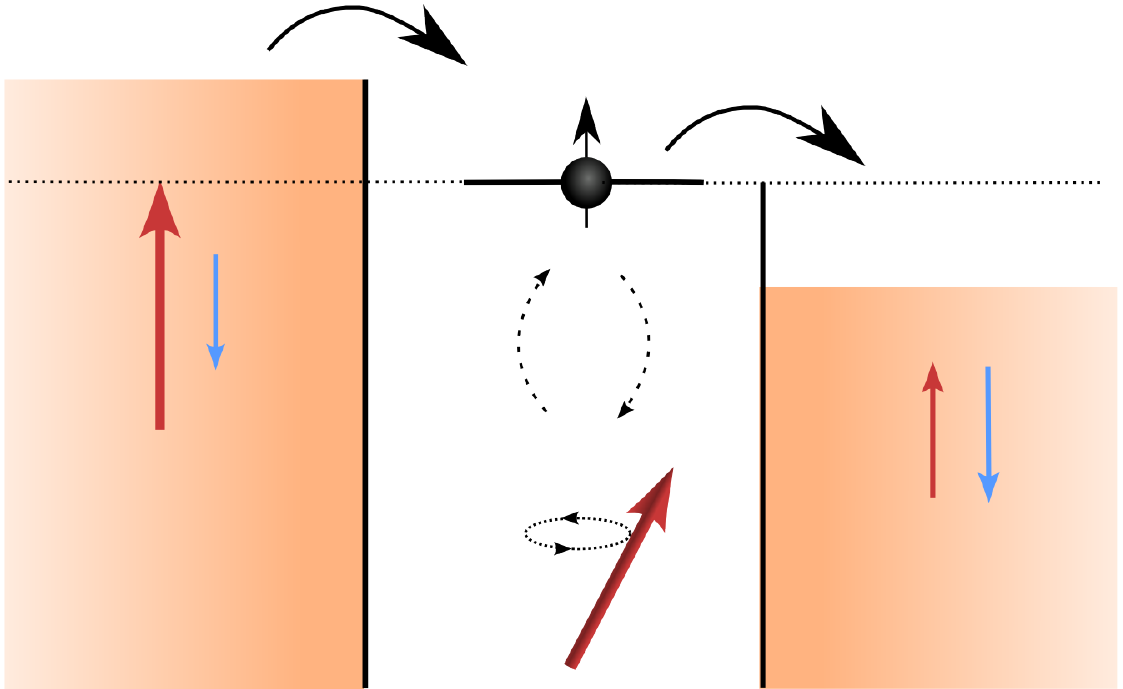}	
	\caption{The system studied in this work consisting of a local magnetic moment coupled to a QD in a tunnel junction between ferro- and non-magnetic leads.}
	\label{system}
\end{figure}

The model system, onto which we apply our developed method, is comprised of a magnetic molecule that is embedded in the tunnel junction between metallic leads. The leads themselves may support spin-polarized currents. Here, the magnetic molecule consists of two components, namely, a quantum dot (QD) level and a localized magnetic moment, that interact via exchange. The QD level is tunnel coupled to the leads. Hence, the current flowing through the metal-QD-metal complex is expected to probe the presence of the localized magnetic moment, and, vice versa, the localized magnetic moment is expected to depend on the current. Taking this observation as an initial condition for our studies, we construct a calculation scheme in which the dynamics of the localized magnetic moment is described by a generalized version of the Landau-Lifshitz-Gilbert equation \cite{fransson2008,PhysRevB.77.205316,franssonNJP2008,PhysRevLett.108.057204}. The effective spin-spin interactions are mediated by the tunneling current flowing across the junction. The current, on the other hand, depends directly on the presence and dynamics of the localized magnetic moment. We include this dependence by feeding the time-evolution of the spin dynamics into the current, which causes the current-dependent temporal spin fluctuations to generate signatures back into the current.

The effective spin model derived in Sec. \ref{sec-method}, depends only on the parameters included in our microscopical model -- there are no \emph{ad hoc} contributions in the description in addition to the basic model. However, within the realms of the model, there is a current mediated spin-spin interaction generated in the effective spin model, which describes interactions between the spin at time $t$ and time $t'$. Hence, although there is only one spin in the system, it is still justified to introduce the concept of spin-spin interaction since the spin at different times can be regarded as different spins.

Separation of the magnetic molecule into a QD level and a localized magnetic moment is justified for, e.g., M-phthalocyanines and M-porphyrins. In these compounds, the transition metal $d$-levels, which are deeply localized, constitute the localized magnetic moment. The $s$- and $p$-orbitals in the ligands, on the other hand, generate the spectral intensity at the highest occupied molecular orbital (HOMO) and lowest unoccupied molecular orbital (LUMO) levels which may be considered as the QD level(s) in our model.

Previously, Bode et al. \cite{PhysRevB.85.115440} performed a similar theoretical treatment of this problem using a non-equilibrium Born-Oppenheimer approximation. Here, however, we go beyond the adiabatic limit and extend the model to the non-Markovian regime in order to treat memory effects and its impact on the exchange interaction. Hence, the interaction fields in the spin equation of motion are not only time-dependent but also dependent on their time-evolutions. A major difference with this formulation is that all retardation effects are included in the time integration of the interaction fields, and it is, therefore, not meaningful to discuss quantities such as Gilbert damping since such parameters are defined in the adiabatic limit.

In general, there also exist stochastic field acting on the localized spin as a result of its interaction with the surrounding electrons. Here, we have chosen to omit the action of these fields, despite their importance for a full description of the physics \cite{Arrachea:2015aa}. However, since we consider the physics in the wide band limit, these electronically induced stochastic fields are of Gaussian white noise character with no voltage bias dependence; see Appendix \ref{app-rv}. The stochastic field in this limit will, therefore, merely play the role of a structureless thermal noise field. As the main focus of our work is on the dynamics of the localized magnetic moment and the exchange interactions, we notice that our results are valid whenever the energies of the interactions are larger than the corresponding energies of these thermal noise fields. Adding a Langevin term, which arises from the quantum fluctuation in the spin action \cite{franssonNJP2008}, into the spin equation of motion could be an interesting extension of the model used in this work, which would be the objective for a separate study.

The paper is organized as follows. In Sec. \ref{sec-method} we discuss the basic set-up of the formalism we employ in this study. After defining the model for the magnetic molecular QD, we derive the equations for the spin moment and the tunneling current. Numerical results from these equations are presented in Sec. \ref{sec-results} and we summarize and conclude the paper in Sec. \ref{sec-summary}.

\section{Method}
\label{sec-method}
To be specific, we consider a magnetic molecule embedded in a tunnel junction between metallic leads that may support spin-polarized currents, see Fig. \ref{system} for reference. The magnetic molecule comprises a localized magnetic moment ${\bf S}$ coupled via exchange to the highest occupied molecular orbital (HOMO) or lowest unoccupied molecular orbital (LUMO) level, henceforth referred to as the QD level.
We define our system Hamiltonian as
\begin{equation}
{\cal H}
	=
	{\cal H_{{\rm \chi}}}+{\cal H_{{\rm T}}}+{\cal H_{{\rm QD}}}+{\cal H_{{\rm S}}}
	.
\end{equation}
Here, ${\cal H}_{\chi}=\sum_{\bfk\in\chi,\sigma}(\varepsilon_{\bfk\chi\sigma}-\mu_{\chi})c_{\bfk\chi\sigma}^{\text{\ensuremath{\dagger}}}c_{\bfk\chi\sigma}$ is the Hamiltonian for the lead $\chi = L/R$, where $c_{\bfk\chi\sigma}^{\dagger}$ ($c_{\bfk\chi\sigma}$) creates (annihilates) an electron in the lead with energy $\varepsilon_{\bfk\chi\sigma}$, momentum \textbf{k} and spin $\sigma=\up,\down$. We have introduced the chemical potential $\mu_{\chi}$ for the leads and the voltage bias across the junction defined as $V = \mu_{L}-\mu_{R}$. Tunneling between the leads and the QD level is described by ${\cal H}_{T}={\cal H}_{TL}+{\cal H}_{TR}$, where ${\cal H}_{T\chi}=T_{\chi}\sum_{\bfk\sigma\in\chi}c_{\mathbf{k\chi}\sigma}^{\dagger}d_{\sigma}+H.c.$. The single-level QD is represented by ${\cal H}_{QD}=\sum_{\sigma}\varepsilon_{\sigma}d_{\sigma}^{\text{\ensuremath{\dagger}}}d_{\sigma}$, where $d_{\sigma}^{\dagger}$ ($d_{\sigma}$) creates (annihilates) an electron in the QD with energy $\varepsilon_{\sigma} = \varepsilon_{0} + g\mu_{B} B \sigma^{z}_{\sigma\sigma}/2$ and spin $\sigma$. We include the Zeeman split due to the external magnetic field $\bfB=B\hat{\bf z}$ where g is the gyromagnetic ratio and $\mu_{B}$ the Bohr magneton. The local spin is described by ${\cal H}_{\rm S}=-g\mu_B\bfS\cdot\bfB-v\mathbf{s}\cdot\mathbf{S}$, where $v$ is the interacting rate between the local spin and the electron spin \textbf{$\mathbf{s}=\sum_{\sigma\sigma'}d_{\sigma}^{\dagger}\boldsymbol{\sigma}_{\sigma\sigma'}d_{\sigma'}/2$}, whereas $\boldsymbol{\sigma}_{\sigma\sigma'}$ is the vector of Pauli matrices.

\subsection{Equation of motion of the local magnetic moment}
Using the methods in, e.g., Refs. \cite{fransson2008,PhysRevB.77.205316,franssonNJP2008,PhysRevLett.108.057204,PhysRevLett.113.257201} and Appendix \ref{app-rv}, we derive an effective spin model for the localized magnetic moment ${\bf S}(t)$ from which we obtain the equation of motion
\begin{eqnarray}
\dot{\mathbf{S}}(t)=
	-g\mu_{B}\mathbf{S}(t)\times\mathbf{B}^\text{eff}(t)
	+\frac{1}{e} \mathbf{S}(t)\times\int\mathbb{J}(t,t')\cdot\mathbf{S}(t')dt'.
\label{spinequationofmotion}
\end{eqnarray}
Here, in order to arrive at this result we have neglected longitudinal spin fluctuations ($\partial_{t}|\mathbf{S}|=0$) and rapid quantum fluctuations. The effective magnetic field is defined as
\begin{align}
\mathbf{B}^\text{eff}(t)=&
	\textbf{B}
	+
	\frac{1}{eg\mu_{B}}\int\boldsymbol{\epsilon}\textbf{j}(t,t')dt',
\end{align}
where $\mathbf{B}$ is the external magnetic field while the second term provides the internal magnetic field due to the electron flow, where
\begin{align}
\boldsymbol{\epsilon}\textbf{j}(t,t')=&
	ie\boldsymbol{\epsilon}v\theta(t-t')\av{\com{\boldsymbol{s}^{(0)}(t)}{\boldsymbol{s}(t')}},
\end{align}
Here, $\boldsymbol{\epsilon}={\rm diag}\lbrace \varepsilon_\uparrow\ \varepsilon_\downarrow\rbrace$ and the charge $\mathbf{s}^{(0)}=\sum_{\sigma\sigma'}d_{\sigma}^{\dagger}\sigma^0_{\sigma\sigma'}d_{\sigma'}/2$, where $\sigma^0$ is the identity matrix. This two-electron Green function (GF) is approximated by a decoupling into single electron GFs according to
\begin{eqnarray}
\boldsymbol{\epsilon}\mathbf{j}(t,t')&\approx&
	iev\theta(t-t'){\rm sp}\boldsymbol{\epsilon}
	\Bigl(
		\mathbf{G}^{<}(t',t)\mathbf{\boldsymbol{\sigma}G}^{>}(t,t')
\nonumber\\&&	
		-\mathbf{G}^{>}(t',t)\mathbf{\boldsymbol{\sigma}G}^{<}(t,t')
	\Bigr),
\label{currentMF}
\end{eqnarray}
where $\mathbf{G}^{</>}(t',t)$ is the lesser/greater matrix GF of the QD defined by ${\bf G}^<(t,t')=\{i\langle c^\dagger_{\sigma'}(t')c_\sigma(t)\rangle\}_{\sigma\sigma'}$ and ${\bf G}^>(t,t')=\{(-i)\langle c_\sigma(t)c^\dagger_{\sigma'}(t')\rangle\}_{\sigma\sigma'}$. In Eq. (\ref{currentMF}) ${\rm sp}$ denotes the trace over spin 1/2 space.

The current $\mathbb{J}(t,t')=i2ev^2\theta(t-t')\av{\com{\bfs(t)}{\bfs(t')}}$ is the electron spin-spin correlation function which mediates the interactions between the localized magnetic moment at times $t$ and $t'$. As with the internal magnetic field, we decouple this two-electron GF according to
\begin{align}
\mathbb{J}(t,t')\approx&
	\frac{ie}{2}v^{2}\theta(t-t'){\rm sp}\boldsymbol{\sigma}
		\Bigl(
			\mathbf{G}^{<}(t',t)\mathbf{\boldsymbol{\sigma}G}^{>}(t,t')
\nonumber\\&
			-\mathbf{G}^{>}(t',t)\mathbf{\boldsymbol{\sigma}G}^{<}(t,t')
		\Bigr).
\end{align} 
This current mediated interaction can be decomposed into an isotropic Heisenberg, $J_H$, interaction and the anisotropic Dzyaloshinski-Moriya (DM), ${\bf D}$, and Ising, $\mathbb{I}$, interactions. This can be seen from the product ${\bf S}\cdot\mathbb{J}\cdot{\bf S}$, which is the corresponding contribution in the effective spin model \cite{PhysRevLett.113.257201} to ${\bf S}(t)\times\mathbb{J}(t,t')\cdot{\bf S}(t')$ in the spin equation of motion. Using the general partitioning ${\bf G}=G_0\sigma^0+{\bf G}_1\cdot\boldsymbol{\sigma}$, where $G_0$ and ${\bf G}_1$ describes the electronic charge and spin, it is straight forward to see that
\begin{align}
{\rm sp}&
	\bfS\cdot\bfsigma\bfG\bfsigma\bfG\cdot\bfS
\nonumber\\=&
	{\rm sp}\bfS\cdot\bfsigma(G_0\sigma^0+\bfG_1\cdot\bfsigma)\bfsigma(G_0\sigma^0+\bfG_1\cdot\bfsigma)\cdot\bfS
\nonumber\\=&
	{\rm sp}
	\Bigl(
		\bfS\cdot\bfG_1
		+
		[
			\bfS G_0+i\bfS\times\bfG_1
		]
		\cdot\bfsigma
	\Bigr)
	\Bigl(
		\bfG_1\cdot\bfS
\nonumber\\&\hspace{3cm}
		+
		[
			\bfG_0\bfS-i\bfG_1\times\bfS
		]
		\cdot\bfsigma
	\Bigr),
\end{align}
where we have used the identity $(\bfA\cdot\bfsigma)(\bfB\cdot\bfsigma)=\bfA\cdot\bfB+i[\bfA\times\bfB]\cdot\bfsigma$. As the Pauli matrices are traceless, the above expression reduces to
\begin{align}
	2
	\Bigl(
		\bfS\cdot(\bfG_1\bfG_1)\cdot\bfS
		+
		[
			\bfS G_0+i\bfS\times\bfG_1
		]
		\cdot
		[
			G_0\bfS-i\bfG_1\times\bfS
		]
	\Bigr).
\end{align}
After a little more algebra we obtain the Heisenberg ($J_{H}$), anisotropic Ising ($\mathbb{I}$) and anisotropic Dzyaloshinsky-Moriya (\textbf{D}) interactions
\begin{subequations}
\label{eq-JID}
\begin{eqnarray}
J_{H}(t,t') & = & iev^{2}\theta(t-t')\left(G_{0}^{<}(t',t)G_{0}^{>}(t,t')\right.\nonumber \\
&  & \left.-G_{0}^{>}(t',t)G_{0}^{<}(t,t')-\mathbf{G}_{1}^{<}(t',t)\cdot\mathbf{G}_{1}^{>}(t,t')\right.\nonumber \\
&  & \left.+\mathbf{G}_{1}^{>}(t',t)\cdot\mathbf{G}_{1}^{<}(t,t')\right),
\label{Heisenberg}
\end{eqnarray}
\begin{eqnarray}
\mathbb{I}(t,t') & = & iev^{2}\theta(t-t')\left(\mathbf{G}_{1}^{<}(t',t)\mathbf{G}_{1}^{>}(t,t')\right.\nonumber \\
&  & \left.-\mathbf{G}_{1}^{>}(t',t)\mathbf{G}_{1}^{<}(t,t')+\left[\mathbf{G}_{1}^{<}(t',t)\mathbf{G}_{1}^{>}(t,t')\right.\right.\nonumber \\
&  & \left.\left.-\mathbf{G}_{1}^{>}(t',t)\mathbf{G}_{1}^{<}(t,t')\right]^{t}\right),
\label{Ising}
\end{eqnarray}
\begin{eqnarray}
\mathbf{D}(t,t') & = & -ev^{2}\theta(t-t')\left(G_{0}^{<}(t',t)\mathbf{G}_{1}^{>}(t,t')\right.\nonumber \\
&  & \left.-G_{0}^{>}(t',t)\mathbf{G}_{1}^{<}(t,t')-\mathbf{G}_{1}^{<}(t',t)G_{0}^{>}(t,t')\right.\nonumber \\
&  & \left.+\mathbf{G}_{1}^{>}(t',t)G_{0}^{<}(t,t')\right).
\label{DM}
\end{eqnarray}
\end{subequations}
This leads to that we can partition the current mediated spin-spin interaction in the spin equation of motion into
\begin{align}
\mathbf{S}(t)\times\mathbb{J}(t,t')\cdot\mathbf{S}(t')=& 
	J_{H}(t,t')\mathbf{S}(t)\times\mathbf{S}(t')
\nonumber\\&
	+\mathbf{S}(t)\times\mathbb{I}(t,t')\cdot\mathbf{S}(t')
\nonumber\\&
	-\mathbf{S}(t)\times\mathbf{D}(t,t')\times\mathbf{S}(t').
\end{align}

In absence of spin-dependence in the QD GF, that is, for $\bfG_1=0$, it is clear that only the Heisenberg interaction $J_H$ remains, since both $\mathbb{I}$ and $\bfD$ explicitly depend on $\bfG_1$. There are different sources that generates a finite $\bfG_1$, e.g., spin injection from the leads, Zeeman split QD level, but also the interaction with the localized magnetic moment gives an essential contribution. In this paper, we include effects from all three sources.

The spin equation of motion derived here goes far beyond the LLG equation as it includes all retardation effects under the time-integration, something which is essentially missing in the LLG equation except for the static exchange interaction and Gilbert damping. Starting from Eq. \ref{spinequationofmotion} and restricting to the adiabatic limit it is possible to derive the conventional LLG equation, see Ref. \cite{PhysRevLett.108.057204}. For clarity this is also done in the Appendix \ref{app-LLGderivation}. This also implies that Eq. \ref{spinequationofmotion} includes the important Gilbert damping and spin-transfer torque, as discussed in Ref. \cite{PhysRevB.85.115440} and \cite{Ralph20081190}. Higher order retardation effects (dissipation, moment of inertia, etc) are included in the time-integral $\int\mathbb{J}(t,t')\cdot\mathbf{S}(t')dt'$.

\subsection{Quantum dot GF}

\subsubsection{Bare quantum dot Green function}
\begin{figure}[t]
	\centering
	\includegraphics[width=\columnwidth]{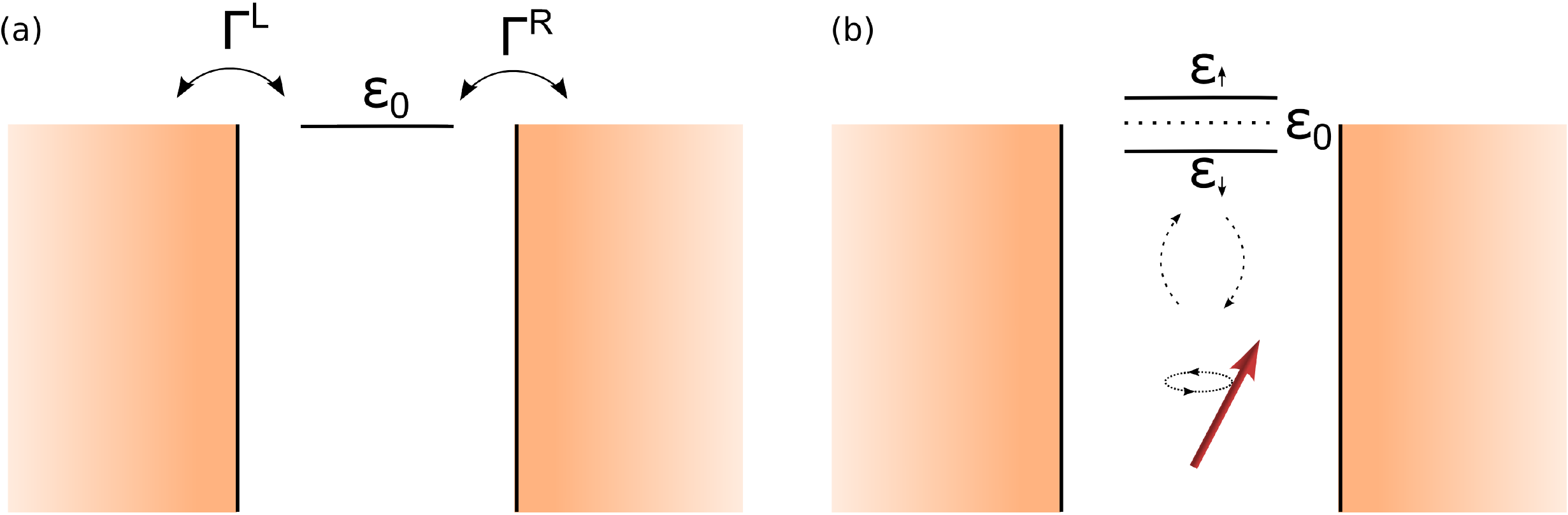}	
	\caption{Sketch of the system without (a) and with (b) a local magnetic moment and coupled to the leads. In the latter case the interactions with the spin moment induce an effective Zeeman split.}
	\label{simplesystem}
\end{figure}

Next, we derive the GF for the QD, which is defined as ${\bf G}(t,t')=\{\eqgr{c^\dagger_{\sigma'}(t')}{c_\sigma(t)}\}_{\sigma\sigma'}$ where T is the contour-ordering operator. We introduce a bare GF $g_{\sigma}(t,t')$ as the solution to the equation
\begin{align}
(i\partial_{t}-\varepsilon_{\sigma})g_{\sigma}(t,t')=&
	\delta(t-t')+\int \Sigma_{\sigma}(t,\tau)g_{\sigma}(\tau,t')d\tau
	.
\end{align}
The bare GF then describes the electronic structure of the QD when coupled to the leads through the self-energy $\Sigma_\sigma(t,t')=\sum_\chi\sum_{\bfk\in\chi}|T_\chi|^2g_{\bfk\sigma}(t,t')$, however, without any coupling to the local spin moment, as illustrated in Fig. \ref{simplesystem} (a). Here,
\begin{align}
g_{\bfk\sigma}(t,t')=&
	(-i){\rm T}e^{-i\int_{t'}^t\leade{\bfk}(\tau)d\tau}
\end{align}
is the GF for the lead $\chi$, including the time-dependence imposed by the voltage bias.

The self-energy $\bfSigma=\diag{\Sigma_\up\ \Sigma_\down}{}$ is treated in the wide-band limit (WBL), which for the retarded/advanced and lesser/greater forms are given by
\begin{subequations}
\begin{align}
\Sigma_\sigma^{r/a}(t,t')=&
	(\mp i)\delta(t-t')\Gamma_\sigma/2
	,
\\
\Sigma_\sigma^{</>}(t,t')=&
	(\pm i)\sum_\chi\Gamma^\chi_\sigma K^{</>}_\chi(t,t'),
\end{align}
\end{subequations}
where $\Gamma_{\sigma}=\sum_\chi\Gamma_\sigma^\chi$ and $\Gamma_\sigma^\chi=2|T_\chi|^2\sum_{\bfk\in\chi}\delta(\omega-\leade{\bfk})$, whereas
\begin{align}
K_{\chi}^{</>}(t,t')=\int f_{\chi}(\pm\omega)e^{-i\omega(t-t')+i\int_{t'}^{t}\mu_{\chi}(\tau)d\tau}\frac{d\omega}{2\pi}
\label{eq:Vinteraction}
	.
\end{align}
Here, $f(\pm\omega)$ is the Fermi function.
The WBL allows to write the retarded/advanced zero GF as 
\begin{equation}
g_{\sigma}^{r/a}(t,t')=(\pm i)\theta(\pm t\mp t')e^{-i(\varepsilon_{\sigma}\mp i\Gamma_{\sigma}/2)(t-t')}
.
\end{equation}

By defining the coupling parameters $\Gamma^{\chi}_{0}=\sum_{\sigma}\Gamma^{\chi}_{\sigma}$ and $\boldsymbol{\Gamma}^{\chi}_1=\sum_{\sigma}\sigma_{\sigma\sigma}^{z}\Gamma^{\chi}_{\sigma}\mathbf{\hat{z}}$ and introducing the spin-polarization in the leads $p_{\chi}\in\left[ -1,1\right]$, such that $\Gamma^{\chi}_{\sigma}=\Gamma_{0}^{\chi}(1+\sigma_{\sigma\sigma}^{z}p_{\chi})/2$, we can write $\boldsymbol{\Gamma}^{\chi}_1=p_{\chi}\Gamma^{\chi}_{0}\mathbf{\hat{z}}$. With this notation we can introduce the coupling matrix $\bfGamma=\Gamma_0\sigma^0+\bfGamma_1\cdot\bfsigma$, where $\Gamma_0=\sum_\chi\Gamma^\chi_0$ and $\bfGamma_1=\sum_\chi\bfGamma^\chi_1$.
Analogously, we write the retarded/advanced and lesser/greater self-energies as $\bfSigma^{r/a}=\Sigma_0^{r/a}\sigma^0+\bfSigma_1^{r/a}\cdot\bfsigma$ and $\bfSigma^{</>}(t,t')=\Sigma_0^{</>}\sigma^0+\bfSigma_1^{</>}\cdot\bfsigma$, where
\begin{subequations}
\begin{align}
\Sigma_{0}^{r/a}(t,t')=&
	(\pm i)\delta(t-t')\Gamma_{0}/2,
\\
\bfSigma_{1}^{r/a}(t,t')=&
	(\pm i)\delta(t-t')\bfGamma_1/2,
\\
\Sigma_{0}^{</>}(t,t')=&
	(\pm i)\sum_{\chi}\Gamma_{0}^{\chi}K_{\chi}^{</>}(t,t'),
\\
\bfSigma_{1}^{</>}(t,t')=&
	(\pm i)\sum_{\chi}\bfGamma_1^\chi K_{\chi}^{</>}(t,t')
	.
\end{align}
\end{subequations}

Using this notation we partion the bare GF in terms of its charge and magnetic components according to
$\bfg=g_0\sigma^0+\bfsigma\cdot\bfg_1$. The retarded/advanced form of $\bfg$ can then be written
\begin{subequations}
\begin{align}
g_{0}^{r/a}(t,t')=&
	(\pm i)\theta(\pm t\mp t')\sum_{\sigma}e^{-i(\varepsilon_{\sigma}\mp i\Gamma_{\sigma}/2)(t-t')}/2
	,
\\
\bfg_{1}^{r/a}(t,t')=&
	(\pm i)\theta(\pm t\mp t')\sum_{\sigma}\sigma_{\sigma\sigma}^{z}e^{-i(\varepsilon_{\sigma}\mp i\Gamma_{\sigma}/2)(t-t')}\mathbf{\hat{z}}/2
	.
\end{align}
\end{subequations}
Analogously, the lesser/greater forms of $\bfg$ are given by
\begin{align}
\mathbf{g}^{</>}(t,t')\equiv&
	\int\mathbf{g}^{r}(t,\tau)\boldsymbol{\Sigma}^{</>}(\tau,\tau')\mathbf{g}^{a}(\tau',t')d\tau d\tau'
\nonumber\\=&
	g_{0}^{</>}(t,t')\sigma_{0}+\bfsigma\cdot\bfg_{1}^{</>}(t,t')
	,
\end{align}
where (time-dependence of the propagators in the integrands is suppressed)
\begin{subequations}
\begin{align}
g_{0}^{</>}(t,t')=&
	\int\left(g_{0}^{r}\Sigma_{0}^{</>}g_{0}^{a}+\bfg_{1}^{r}\Sigma_{0}^{</>}\cdot\bfg_{1}^{a}\right.
\nonumber\\&
	\left. +g_{0}^{r}\bfSigma_{1}^{</>}\cdot\bfg_{1}^{a}+\bfg_{1}^{r}\cdot\bfSigma_{1}^{</>}g_{0}^{a}\right)d\tau d\tau',
\\
\mathbf{g}_{1}^{</>}(t,t')=&
	\int\left(g_{0}^{r}\bfSigma_{1}^{</>}g_{0}^{a}+\bfg_{1}^{r}\cdot\bfSigma_{1}^{</>}\bfg_{1}^{a}\right.
\nonumber\\&
	\left. +g_{0}^{r}\Sigma_{0}^{</>}\bfg_{1}^{a}+\bfg_{1}^{r}\Sigma_{0}^{</>}g_{0}^{a}\right)d\tau d\tau'.
\end{align}
\end{subequations}

\subsubsection{Dressed quantum dot Green function}
The next step is to include the interactions with the local magnetic moment into the description. We achieve this goal by defining the dressed QD GF as the first order expansion in terms of the local moment, that is,
\begin{align}
\mathbf{G}(t,t')=&
	\mathbf{g}(t,t') + \delta\mathbf{G}(t,t')
\nonumber\\=&
	\mathbf{g}(t,t')-v\oint_C\mathbf{g}(t,\tau)\left\langle \mathbf{S}(\tau)\right\rangle \mathbf{\cdot\boldsymbol{\sigma}g}(\tau,t')d\tau.
\label{eq:Dressed Greens function}
\end{align}
where $\mathbf{g}$ is the bare GF and $\delta\mathbf{G}$ is the correction from the interactions with the local magnetic moment. As above, we write
$\bfG=\bfG_0\sigma^0+\bfsigma\cdot\bfG_1$, where $G_{0}=g_{0}+\delta G_{0}$ and $\bfG_{1}=\bfg_1+\delta\bfG_1$, whereas the corrections are given by
\begin{subequations}
\begin{align}
\delta G_{0}(t,t')=&
	-v\oint_C
		\Bigl(
			g_{0}\av{\bfS}\cdot\bfg_1
\nonumber\\&
			+
			\bfg_1\cdot\av{\bfS}g_0
			+
			i[\bfg_1\times\av{\bfS}]\cdot\bfg_1
		\Bigr)
	d\tau,
\\
\delta\bfG_1(t,t')=&
	-v\oint_C
		\Bigl(
			g_0\av{\bfS}g_0
			+
			i[\bfg_1\times\av{\bfS}]g_0
\nonumber\\&
			+
			ig_0[\av{\bfS}\times\bfg_1]
			+
			i[\bfg_1\times\av{\bfS}]\times\bfg_1
		\Bigr)
	d\tau
	.
\end{align}
\end{subequations}
We refer to Appendix \ref{app-NEGF} for details about the lesser/greater forms of the charge and magnetic components of $\delta\bfG$.

It should be noticed that the presence of the local spin moment gives rise to a spin-polarization of the QD level due to the local exchange interaction, see Fig. \ref{simplesystem} (b) for an illustration. The effect is particularly strong whenever there is an intrinsic spin-polarization in either the leads and/or the QD, in which case $\bfg_1\neq0$. Then, the local spin moment affects the properties of both the charge and magnetic structure of the QD. Nevertheless, even for spin-degenerate leads and QD, that is, for $\bfg_1\equiv0$, the QD level acquires a finite spin-dependence. This is legible in the expression for $\delta\bfG_1$, where the first term only depends on the magnetic properties of the local spin moment and the charge density in the QD.
Thus, by calculating the electronic structure in the QD as function of the local spin moment opens for tracing signatures of the local spin dynamics in the properties of the QD.

\subsection{Current}
The properties of the QD are probed by means of the electron currents flowing through the system. In this way, the goal is to pick up signatures of the spin dynamics in the transport properties as these should influence the electronic structure of the QD. The electron currents can be decomposed into charge and spin currents, $I^C$ and $I^S$, respectively. Here, we calculate the currents flowing through the left interface between the leads and the QD. Accordingly, we define
\begin{subequations}
\begin{align}
I^C_L(t)=&
	-e\dt\sum_{\bfk\sigma\in L}\av{n_{\bfk\sigma}}=ie{\rm sp}\dt\sum_\bfk\bfG_\bfk^<(t,t)
	,
\\
I^S_L(t)=&
	-e\dt\sum_{\bfk\sigma\sigma'\in L}\av{\cdagger{\bfk}\bfsigma_{\sigma\sigma'}\cs{\bfk\sigma'}}=ie{\rm sp}\bfsigma\dt\sum_\bfk\bfG^<_\bfk(t,t)
	.
\end{align}
\end{subequations}
Using standard methods we can write the charge current as
\begin{eqnarray}
I_{L}^{C}(t) & = & -\frac{2e}{\hbar}\operatorname{sp}\operatorname{Im}\boldsymbol{\Gamma^{L}}\int_{-\infty}^{t}\left(K_{L}^{>}(t,t')\mathbf{G}^{<}(t',t) \right.\nonumber \\
&  & \left. +K_{L}^{<}(t,t')\mathbf{G}^{>}(t',t)\right)dt'\label{chargecurrent}
	.
\end{eqnarray}
Following the same route as initiated above, we partition the current into a spin-independent and spin-dependent part according to $I_{L}^{C}(t)  =  I_{0}^{C}(t)+I_{1}^{C}(t)$, where
\begin{subequations}
\begin{align}
I_{0}^{C}(t)=&
	\frac{4e}{\hbar}
	\Gamma_0^L
	\im
	\int_{-\infty}^t
		\Bigl(
			K_L^>G_0^<
			+
			K_L^<G_0^>
		\Bigr)
	dt'
		,
\label{chargecurrent0}
\\
I_{1}^{C}(t)=&
	-\frac{4e}{\hbar}
	\bfGamma_1^L\cdot
	\im
	\int_{-\infty}^t
		\Bigl(
			K_L^>\bfG_1^<
			+
			K_L^<\bfG_1^>
		\Bigr)
	dt'
		.
\label{chargecurrent1}
\end{align}
\end{subequations}

Analogously to the charge current, we write the spin current as
\begin{align}
\mathbf{I_{L}^{S}}(t)=&
	-\frac{2e}{\hbar}\operatorname{sp}\operatorname{Im}\boldsymbol{\sigma}\boldsymbol{\Gamma^{L}}\int_{-\infty}^{t}\left(K_{L}^{>}(t,t')\mathbf{G}^{<}(t',t)\right.
\nonumber \\&
		\left. +K_{L}^{<}(t,t')\mathbf{G}^{>}(t',t)\right)dt'
		,
\label{spincurrent}
\end{align}
where $\mathbf{I_{L}^{S}}(t) = \mathbf{I_{0}^{S}}(t)+\mathbf{I_{1}^{S}}(t)$ and
\begin{subequations}
\label{spincurrent01}
\begin{align}
\mathbf{I_{0}^{S}}(t)=&
	-\frac{4e}{\hbar}
	\Gamma_0^L
	\im
	\int_{-\infty}^t
		\Bigl(
			K_L^>\bfG_1^<
			+
			K_L^<\bfG_1^>
		\Bigr)
	dt'
		,
\label{spincurrent0}
\\
\mathbf{I_{1}^{S}}(t)=&
	-\frac{4e}{\hbar}
	\im
	\int_{-\infty}^t
		\biggl[
			K_L^>
			\Bigl(
				\bfGamma_1^LG_0^<
				+
				i\bfGamma_1^L\times\bfG_1^<
			\Bigr)
\nonumber\\&
			+
			K_L^<
			\Bigr(
				\bfGamma_1^LG_0^>
				+
				i\bfGamma_1^L\times\bfG_1^>
			\Bigr)
		\biggr)
	dt'
		.
\label{spincurrent1}
\end{align}
\end{subequations}

These expressions for the charge and spin currents suggest that any local dynamics that is picked up by the electronic structure of the QD should provide signatures in its transport properties. Next, we analyze the impact of the local dynamics on the transport properties.

\section{Results}
\label{sec-results}
\subsection{Stationary limit}

Before embarking into the full time-dependent properties of the system, we review some of the expected results for the stationary regime in order to provide a benchmark for our calculations.
In the stationary limit all the time-dependences induced from the on-set of the applied voltage bias have decayed which leads to the bare QD GF becomes time local and we can, therefore, study the energetic properties of the QD.
Then, the local magnetic moment, $\av{\bfS}$, can be regarded as a constant spin-polarization and a source for coupling between the spin states, in agreement with Ref. \cite{PhysRevB.87.045426}.
The Fourier transform of the bare QD GF is, therefore, written on the form
\begin{subequations}
\begin{align}
g_0^{r/a}(\omega)=&
	\frac{1}{2}
	\sum_\sigma
		g_\sigma^{r/a}(\omega),
\\
\bfg_1^{r/a}(\omega)=&
	\frac{1}{2}
	\sum_\sigma
		\sigma^z_{\sigma\sigma}
		g_\sigma^{r/a}(\omega),
		\hat{\bf z},
\end{align}
\end{subequations}
where
\begin{align}
g_\sigma^{r/a}(\omega)=&
	\frac{1}{\omega-\dote{0}\pm i\Gamma_\sigma/2}
,
\end{align}
and the self-energies become 
\begin{subequations}
\begin{align}
\Sigma_0^{</>}(\omega) =&
	(\pm i)\sum_\chi\Gamma_0^\chi f_\chi(\pm\omega),
\\
\bfSigma_1^{</>}(\omega) = &
	(\pm i)\sum_\chi\bfGamma_1^\chi f_\chi(\pm\omega),
\end{align}
\end{subequations}
since $K_{\chi}^{</>}(\omega)= f(\pm\omega)$ in the stationary limit.
%
%
In the stationary limit the interaction parameters, moreover, simplify in the limit $\epsilon \rightarrow 0$ to

\begin{subequations}
	\begin{align}
	J^{(H)}=&
	-v^{2}\int\frac{1}{\omega + \epsilon-  \omega'}\left(G_{0}^{<}(\omega)G_{0}^{>}(\omega')-G_{0}^{>}(\omega)G_{0}^{<}(\omega')\right.
	\nonumber\\&
	\left.-\mathbf{G}_{1}^{<}(\omega)\cdot\mathbf{G}_{1}^{>}(\omega')+\mathbf{G}_{1}^{>}(\omega)\cdot\mathbf{G}_{1}^{<}(\omega')\right)\frac{d\omega}{2\pi}\frac{d\omega'}{2\pi},
	\\
	\mathbb{I}=&
	-v^{2}\int\frac{1}{\omega + \epsilon-  \omega'}\left(\mathbf{G}_{1}^{<}(\omega)\mathbf{G}_{1}^{>}(\omega')-\mathbf{G}_{1}^{>}(\omega)\mathbf{G}_{1}^{<}(\omega')\right.
	\nonumber\\&
	\left.+\left[\mathbf{G}_{1}^{<}(\omega)\mathbf{G}_{1}^{>}(\omega')
	-\mathbf{G}_{1}^{>}(\omega)\mathbf{G}_{1}^{<}(\omega')\right]^{t}\right)\frac{d\omega}{2\pi}\frac{d\omega'}{2\pi},
	\\
	\mathbf{D}=&
	\frac{v^{2}}{2}\re\int\left(G_{0}^{<}(\omega + \epsilon)\mathbf{G}_{1}^{>}(\omega)-G_{0}^{>}(\omega + \epsilon)\mathbf{G}_{1}^{<}(\omega)\right.
	\nonumber\\&
	\left.-\mathbf{G}_{1}^{<}(\omega + \epsilon)G_{0}^{>}(\omega)+\mathbf{G}_{1}^{>}(\omega + \epsilon)G_{0}^{<}(\omega)\right)\frac{d\omega}{2\pi}.
	\end{align}
\end{subequations}

\begin{figure}[t]
	\begin{center}
	\includegraphics[width=\columnwidth]{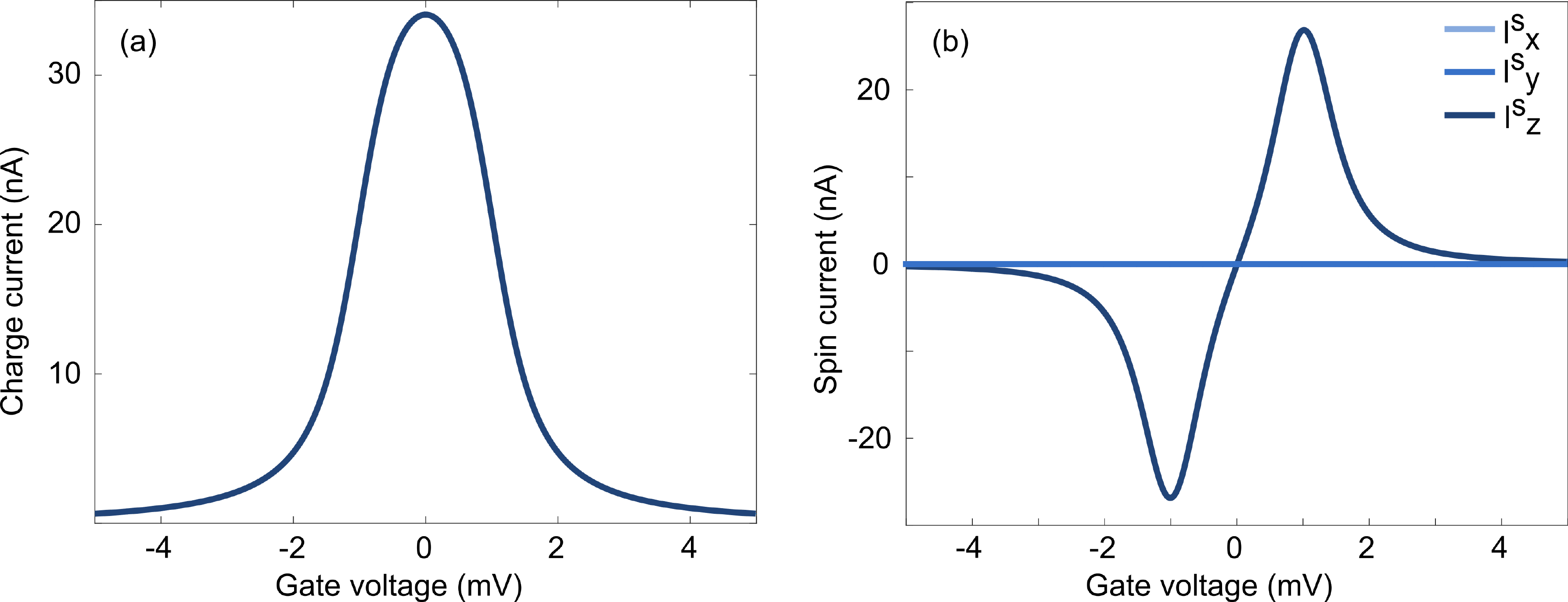}	
	\end{center}
	\caption{
	Charge and spin current for a static local magnetic moment in a tunnel junction.
	(a) Charge current ${\rm I^{C}}$ as function of $\varepsilon_{0}$ and
	(b) spin current ${\rm I^{S}}$ as function of $\varepsilon_{0}$.
	Here, we used $\Gamma_{0} = v = 1$ meV, $T$ = 1 K, $B$ = 1 T, $p_{L}=p_{L}=0$ and $V$ = 2 mV such that $V_{L} = V/2$ and $V_{R} = -V/2$.
	}
	\label{timeindependent}
\end{figure}

\begin{figure}[b]
	\includegraphics[width=\columnwidth]{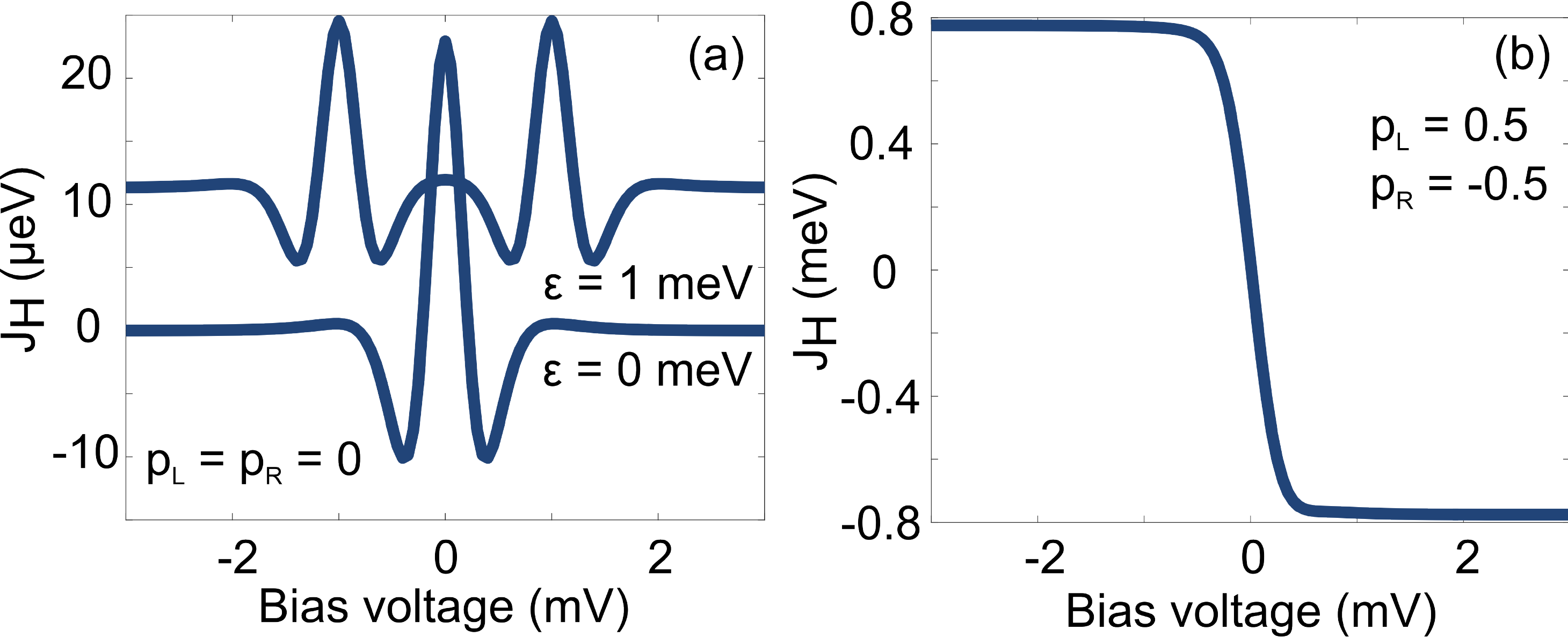}	
	\caption{Heisenberg interaction $J_{H}$ of a static local magnetic moment in a tunnel junction in the z-direction, $\textbf{S} = S_{z}\mathbf{\hat{z}}$. Panel (a) shows the Heisenberg interaction for non-magnetic leads $p_{\chi}=0$ as a function of bias voltage V. Here, the gate voltage is set to $\varepsilon_{0} = 0$ meV and $\varepsilon_{0} = 1$ meV and the plots shifted for clarity (scale is the same). Panel (b) shows the Heisenberg interaction for antiferromagnetic leads, $p_{L}=-p_{R}=0.5$, as a function of bias voltage V. Here, we used $\Gamma_{0} = v = 0.1$ meV, $T$ = 1 K, $B$ = 0 T and $V$ = 2 mV.}
	\label{heisenberg}
\end{figure}

\begin{figure}[t]
	\includegraphics[width=\columnwidth]{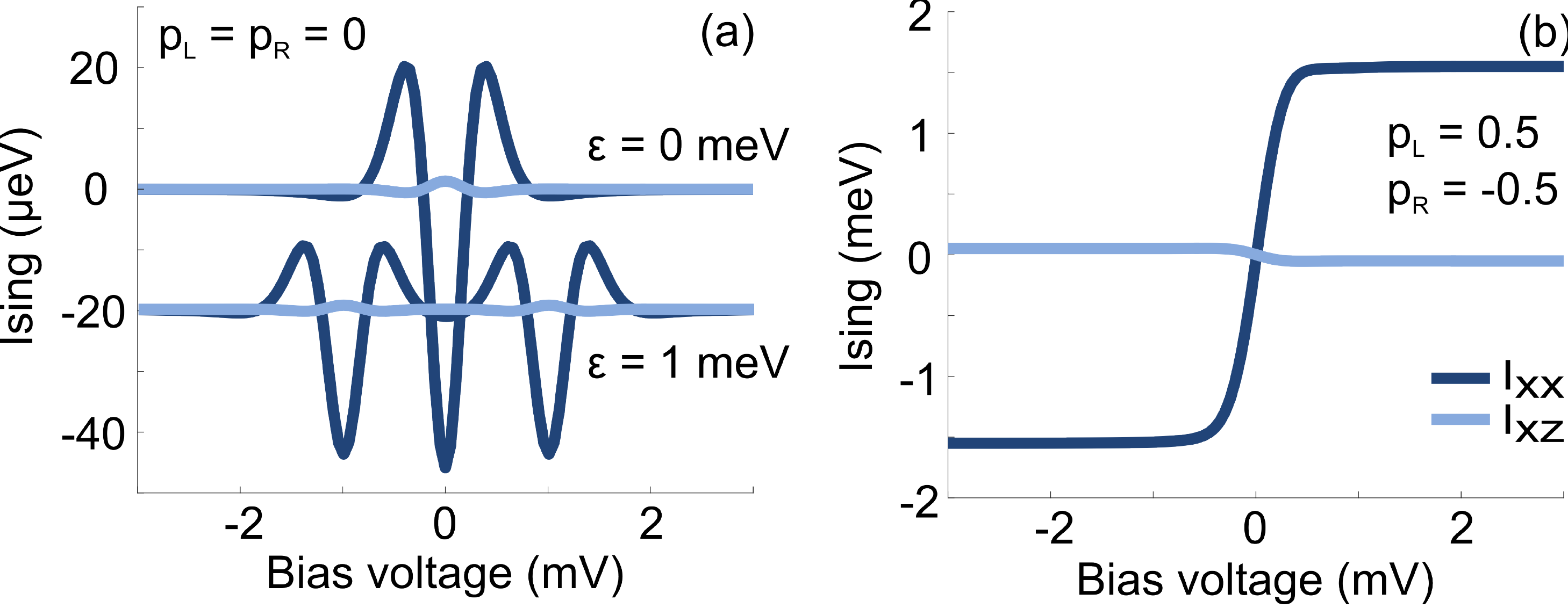}	
	\caption{Ising interaction $\mathbb{I}$ of a static local magnetic moment in a tunnel junction in the z-direction, $\textbf{S} = S_{z}\mathbf{\hat{z}}$. Panel (a) shows the Ising interaction for non-magnetic leads $p_{\chi}=0$ as a function of bias voltage V. Here, the gate voltage is set to $\varepsilon_{0} = 0$ meV and $\varepsilon_{0} = 1$ meV and the plots shifted for clarity (scale is the same). Panel (b) shows the Ising interaction for antiferromagnetic leads, $p_{L}=-p_{R}=0.5$, as a function of bias voltage V. Other parameters as in Fig. \ref{heisenberg}.}
	\label{isingfig}
\end{figure}

\begin{figure}[b]
	\includegraphics[width=\columnwidth]{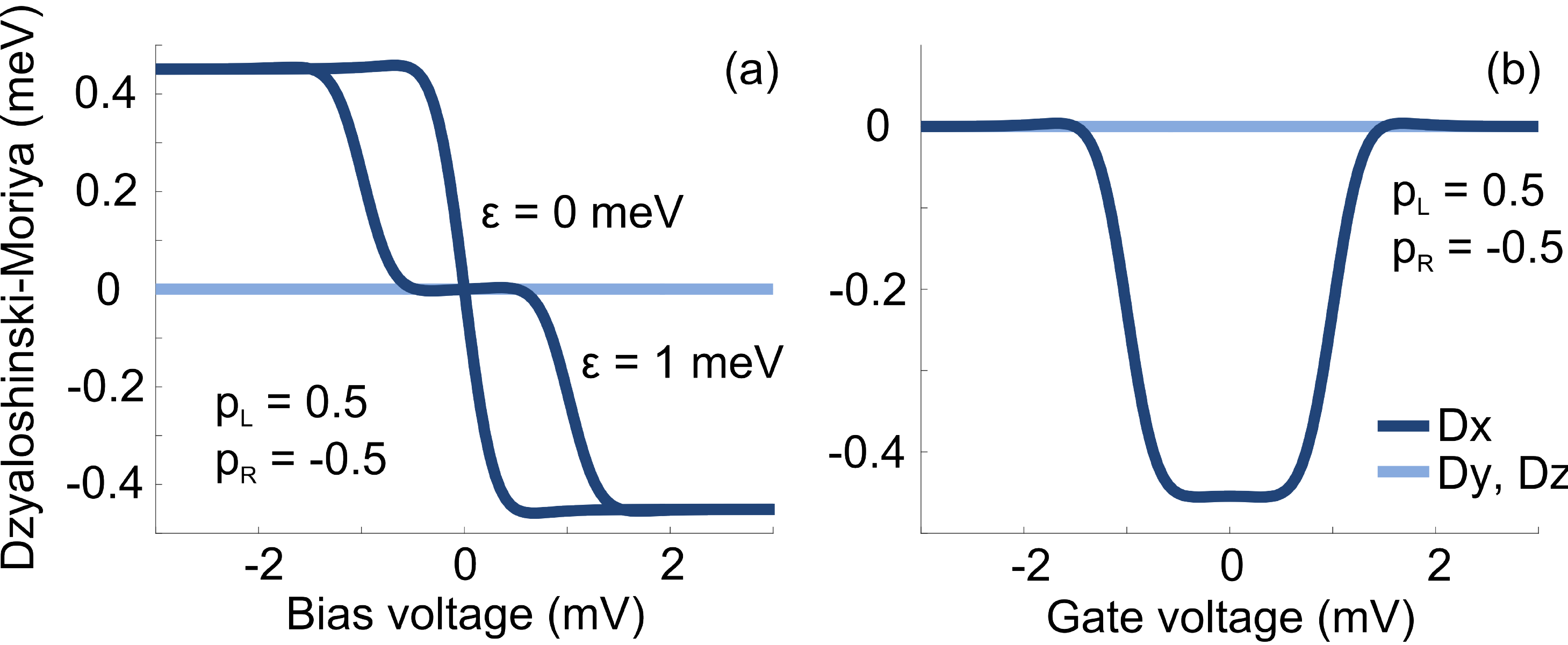}	
	\caption{DM interaction \textbf{D} of a static local magnetic moment in a tunnel junction in the z-direction, $\textbf{S} = S_{z}\mathbf{\hat{z}}$, for antiferromagnetic leads, $p_{L}=-p_{R}=0.5$. Panel (a) shows the DM interaction as a function bias voltage V where the gate voltage is set to $\varepsilon_{0} = 0$ meV and $\varepsilon_{0} = 1$ meV. Panel (b) shows the DM interaction different gate voltage $\varepsilon_{0}$. Other parameters as in Fig. \ref{heisenberg}.}
	\label{DMtimein}
\end{figure}

Considering a symmetric and spin-independent background, i.e. non-magnetic contacts $p_{\chi} = 0$, and a constant local magnetic moment, $\bfS$,
the local spin-polarization gives rise to finite spin currents ${\bf I}^S$ in the system, see Eqs. (\ref{spincurrent}) -- (\ref{spincurrent01}) (note $\bfGamma^S=0$). In Fig. \ref{timeindependent} we plot the calculated (a) charge ($I^C$) and (b) spin current (${\bf I}^S$) as a function of the gate voltage, $V$, for a QD with a bare level at $\dote{0}=0$.
While the charge current behaves as expected for a single-level QD, given by
\begin{align}
I_{L}^{C}=&
	\frac{e}{4\pi\hbar}\Gamma_0^2
	\int\frac{f_{L}(\omega)-f_{R}(\omega)}{(\omega - \epsilon_{0})^2 + (\Gamma_{0}/4)^2}d\omega
,
\label{charge current stationary}
\end{align}
the features in the spin current for gate voltages near zero give a clear indication of the induced spin-polarization from the local spin moment.
Due to the local spin moment induced effective Zeeman split in the QD, as shown in Fig. \ref{simplesystem} (b), the spin current is strongly peaked at $\mu_\chi=\dote{0}$. As can be seen in Fig. \ref{simplesystem}, either one of the spin up- or down channels will be more favorable for the tunneling electrons, thus causing a net spin current in either direction depending on the configuration of the electron level of the leads. This is an important feature as it can be used in order to read out the state of the local spin moment from the spin current.
 

\begin{figure*}[t]
	\centering
	\includegraphics[width=\textwidth]{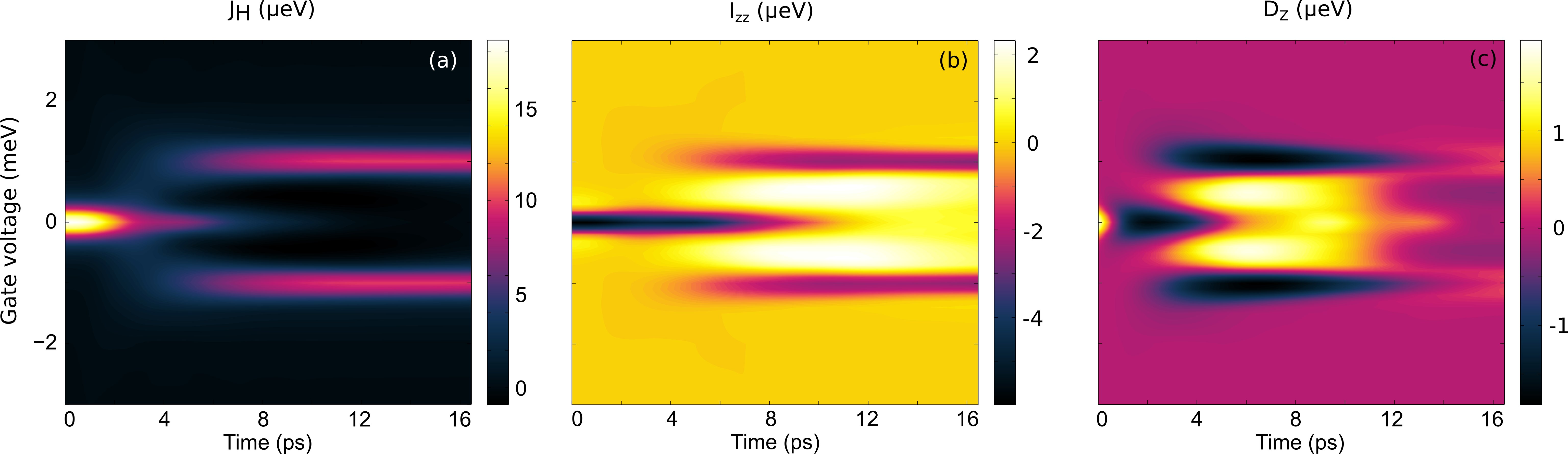}	
	\caption{Time-dependent evolution of the exchange interaction parameters as a function of gate voltage $\varepsilon_{0}$ after an onset of a step-like finite bias voltage of V = 2 mV. Panel (a) shows the strength of the Heisenberg interaction, panel (b) shows the $I_{zz}$ part of the Ising interaction and panel (c) shows the z-component of the DM interaction. Here we used the parameters $\Gamma_{0} = v = 0.1$ meV, $T$ = 1 K, $B$ = 1 T, $p_{L}=p_{L}=0$.}
	\label{timedependentparameters}
\end{figure*}

Regarding the Heisenberg interaction, recalling that, e.g., $G_0^r(\omega)=g_0^r(\omega)=1/(\omega-\dote{0}+i\Gamma_0/4)$ for non-magnetic leads, it can be readily seen that the charge contribution to the Heisenberg exchange is given by
\begin{align}
\frac{2v^2}{\pi}&
	\sum_\chi\Gamma^\chi_0
	\int
		f_\chi(\omega)
	\frac{\omega-\dote{0}}{[(\omega-\dote{0})^2+(\Gamma_0/4)^2]^2}
	d\omega.
\label{eq-HC1}
\end{align}
This suggests a spin-spin interaction which is strongly peaked around $\mu_\chi=\dote{0}$. Similarly, the contribution from the local spin-polarization, $\bfG_1=-vg_0\av{\bfS}g_0$, acquires the form
\begin{align}
-\frac{4v^4}{\pi}&
	|\av{\bfS}|^2
	\sum_\chi\Gamma^\chi_0
	\int
		f_\chi(\omega)
		(\omega-\dote{0})
		\frac{(\omega-\dote{0})^2-(\Gamma_0/4)^2}{[(\omega-\dote{0})^2+(\Gamma_0/4)^2]^4}
	d\omega,
\label{eq-JG1}
\end{align}
which is also strongly peaked at $\mu_\chi=\dote{0}$. However, as the integrand of this component changes sign at $\omega=\dote{0},\dote{0}\pm\Gamma_0/4$, the contribution from the QD spin-polarization goes through local minima at $\dote{0}\pm\Gamma_0/4$ and a local maxima at $\dote{0}$, as a function of the chemical potential $\mu_\chi$.
We therefore expect a competition between the charge and magnetic components which may lead to a change of sign in the Heisenberg interaction, depending both on the properties of the system as well as on the external conditions.
This is illustrated by the computed Heisenberg exchange plotted in Fig. \ref{heisenberg} (a) as a function of the voltage bias for different gate voltages, showing the changing character from negative to positive interaction as the chemical potential $\mu_\chi$ approaches the QD level.
For ferromagnetic leads aligned anti-ferromagnetically in Fig. \ref{heisenberg} (b), $p_{L}=-p_{R}=0.5$, we notice an anisotropic behavior as the sign of the interaction switches with respect to the polarity of the voltage bias. This is in agreement with previous studies of anti-ferromagnetically aligned leads coupled to molecular spins \cite{Misiorny:2013aa}.


The Ising interaction $\mathbb{I}$ essentially behaves in a similar manner, however, this contribution requires a finite spin-polarization ($\bfG_1\neq0$) to become non-vanishing. For non-magnetic leads, $p_{\chi} = 0$, this spin-polarization is provided by the local spin moment and we find that the Ising interaction acquires the form given in Eq. (\ref{eq-JG1}), up to multiplying constants. This is also verified by the numerically computed Ising interaction, shown in Fig. \ref{isingfig} (a) as function of the voltage bias for different gating conditions.
Again, for ferromagnetic leads in anti-ferromagnetic alignment, $p_{L}=-p_{R}=0.5$, there is a switching behavior with respect to the polarity of the voltage bias.

A similar switching behavior appears in the DM interaction \textbf{D}, which is only considered for ferromagnetic leads aligned anti-ferromagnetically, see Fig. \ref{DMtimein} (a), where the DM interaction is plotted as a function of the voltage bias and for different gating conditions. Varying the gate voltage, it can be seen that there is a finite DM interaction only whenever the QD electron level, $\varepsilon_{0}$, lies in the window between the chemical potentials in the leads spanned by the voltage bias. This is understood since the DM interaction results from net current flow interacting with the local spin moment, as it requires simultaneous breaking of time-reversal and inversion symmetries to be finite.

We comment finally on the relevance for calculating the interaction parameters in the stationary limit. This question is justified since the effective spin Hamiltonian in the stationary limit would assume the form
\begin{align}
\Hamil^\text{eff}_S=&
	-J^{(H)}\bfS\cdot\bfS
	-\bfD\cdot\bfS\times\bfS
	-\bfS\cdot\mathbb{I}\cdot\bfS.
\end{align}
Here, one can notice that $\bfS\cdot\bfS=|\bfS|^2$, which is a constant of motion, whereas as $\bfS\times\bfS\equiv0$. Both these identities relies on the fact that the spin $\bfS$ is time-independent in the stationary limit. Actually, only the Ising interaction is physically motivated, providing an anisotropy field on the spin. For collinear spin-polarization in the surrounding system, this contribution reduces to the form $\mathbb{I}_{zz}S_z^2$, which is the ordinary Ising Hamiltonian.

We justify the calculations and analysis of the stationary limit interaction parameters by that we can understand and interpret much of the time-dependent features, discussed in the remainder of this paper, from the results obtained in the stationary limit. In addition, our results also demonstrate that despite the dynamics may be trivial, the fields that mediate the interactions between the dynamical object need not be trivial.

\begin{figure*}[t]
	\centering
	\includegraphics[width=\textwidth]{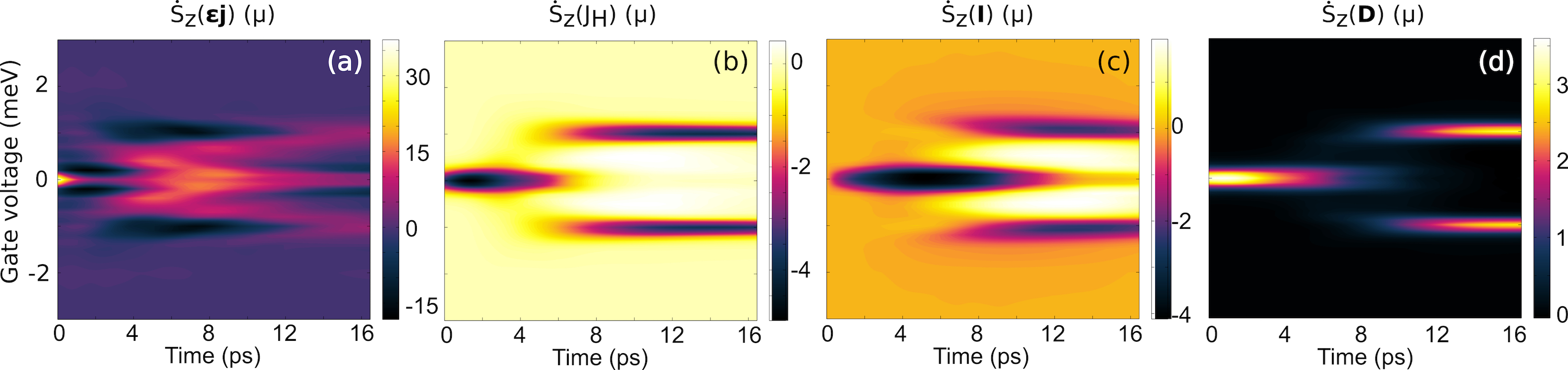}	
	\caption{Contribution to the local magnetic moment equation of motion for different exchange interaction parameters as a function of gate voltage $\varepsilon_{0}$. In (a) the change $\dot{S}_{z}(\mathbb{\epsilon} \textbf{j})$ in z-direction depending on the induced internal magnetic field due to the charge flow. In (b)-(d) the change in the z-direction depending on the Heisenberg, Ising and DM interaction is shown respectively. Other parameters as in Fig. \ref{timedependentparameters}.}
	\label{spinparameters}
\end{figure*}

\begin{figure}[b]
	\centering
	\includegraphics[width=\columnwidth]{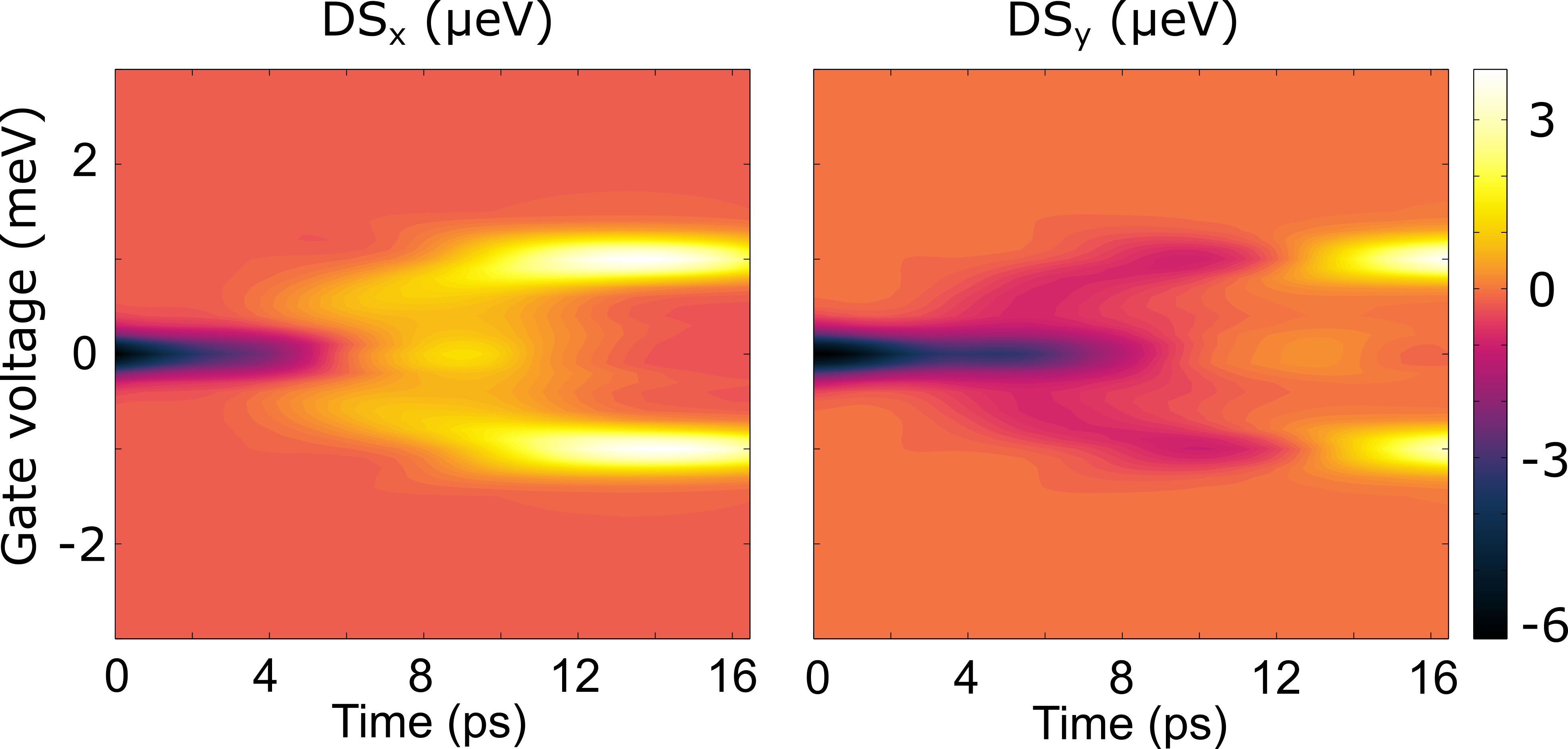}	
	\caption{Time-dependent evolution of the field from the anisotropic DM interaction in x- and y-direction as a function of gate voltage $\varepsilon_{0}$ after an onset of a step-like finite bias voltage of V = 2 mV. Other parameters as in Fig. \ref{timedependentparameters}.}
	\label{DMfield}
\end{figure}

\subsection{Time-dependent exchange interaction}
As we are interested in the transient dynamics, we study the effect of an abrupt on-set of the voltage bias applied as a step-like function $V_{\chi}\theta(t-t_{0})$ symmetrically over the junction such that $V_{L/R} =\pm V/2$. Before the on-set of the voltage bias, the local spin is subject to the static external magnetic field $\mathbf{B}=B\hat{\mathbf{z}}$, giving $S_{x}=S_{xy}\sin\omega_{L}t$, $S_{y}=S_{xy}\cos\omega_{L}t$ and $S_{z}=S_{z}$ where $S_{xy}^2=S_x^2+S_y^2$, whereas $\left|S\right|^{2}=S_{xy}^{2}+S_{z}^{2}$ and $\omega_{L}=g\mu_{B}\left|B\right|$, and we assume an initial polar angle of $\pi/4$.

The time-dependence of the interaction parameters, cf. Eq. (\ref{eq-JID}), has to be calculated as function of the gate voltage and voltage bias at each time-step. In Fig. \ref{timedependentparameters} we plot the time-evolution of the Heisenberg, Ising, and DM interaction parameters as function of the gate voltage, where we integrated over all $t'$, hence, showing $J_{H}(t)$, $I_{zz}(t)$ and $D_{z}(t)$. Considered in this fashion, the plots illustrate the time-evolution of the exchange interaction that would be expected in the adiabatic approximation, that is, $\int\mathbb{J}(t,t')\cdot\bfS(t')dt'\approx\int\mathbb{J}(t,t')dt'\cdot\bfS(t)+\cdots$. In the transient regime, the interaction parameters changes continuously, both due to the changing characteristics of the system and the feed-back through the system from the changing local magnetic moment. In the long time limit, it may be noticed that the interaction strength peaks for all three types of interactions when the QD electron level $\varepsilon_{0}$ is resonant with one of the chemical potentials of the leads $\mu_\chi=eV_{\chi}$. We, hence, retain the properties of the system in the stationary regime.
%

When going beyond the adiabatic approximation, one cannot strictly separate the interaction parameters from the time-evolution of the spin, see for instance Eq. (\ref{spinequationofmotion}). It is then more comprehensible to directly study and analyze each component of the equation of motion.
%
Accordingly, in Fig. \ref{spinparameters} we plot the rates of change in the respective panels
\begin{itemize}
\item[(a)] $\dot{S}_z(t;\boldsymbol{\dote{}}\bfj)=-\bfS(t)\times\int[\boldsymbol{\epsilon}\bfj(t,t')]_zdt'/e$, 
\item[(b)] $\dot{S}_z(t;J^{(H)})=-\bfS(t)\times\int J^{(H)}(t,t')S_z(t')dt'/e$, 
\item[(c)] $\dot{S}_z(t)=-\bfS(t)\times\int \mathbb{I}_{zz}(t,t')S_z(t')dt'/e$, and 
\item[(d)] $\dot{S}_z(t)=-\bfS(t)\times\int\bfD_z(t,t')S_z(t')dt'/e$.
\end{itemize}

The rate of change caused by the current induced magnetic field, $\dot{S}_z(t,\boldsymbol{\dote{}}\bfj)$, shown in Fig. \ref{spinparameters} (a), initially provides a large contribution to the spin dynamics while it tends to zero in the long time limit. This is to be expected since the time variations of the charge current are largest immediately after the on-set of the voltage bias. Far beyond the transient regime initiated by this on-set, the temporal variations in the charge current are much smaller which, therefore, also leads to a smaller induced magnetic field.

Similar behavior appears for the Heisenberg, Ising and DM interactions, shown in Fig. \ref{spinparameters} (b) -- (d), where they initially provide a large contribution to the spin dynamics in the transient regime. Resulting from the time-dependent interaction parameters, there is a finite contribution to the spin dynamics for large time scales in the stationary limit, in agreement with the time-independent solution. The difference, however, is that there is a finite contribution in the stationary limit of the DM-interaction, something not observed in the time-independent solution for non-magnetic leads. The reason of this effect is the time-dependent feature of the DM-fields. In Fig. \ref{DMfield}, the effective field from the DM interaction is shown in the x- and y-components of the vector defined by $(\mathbf{D}\cdot\mathbf{S})(t) = \int\mathbf{D}(t,t')\cdot\mathbf{S}(t')dt'$.
The fields for the Heisenberg and Ising contribution, i.e. $\int J_H(t,t')\mathbf{S}(t')dt'$ and $\int\mathbb{I}(t,t')\times\mathbf{S}(t')dt'$, are similar to the basic parameters shown in Fig. \ref{timedependentparameters}, $\int J_H(t,t')dt'$ and $\int\mathbb{I}(t,t')dt'$.

\begin{figure}[t]
	\centering
	\includegraphics[width=\columnwidth]{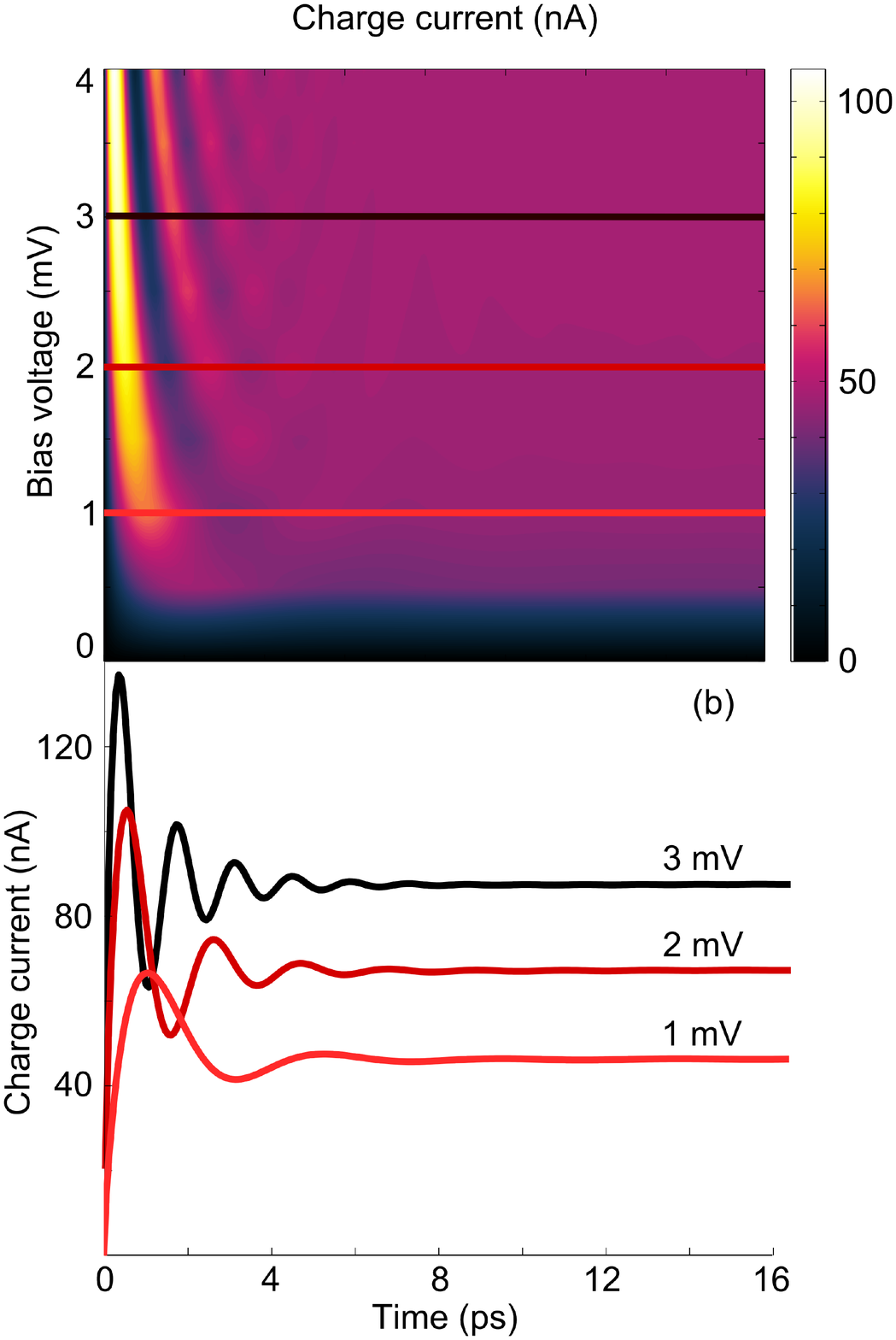}	
	\caption{Charge current as a function of time, ${\rm I^{C}(t)}$, for different step-like bias voltages V. In (a) a contour of the time evolution of the charge current as a function of bias voltage V is shown. In (b) the cuts in (a) are shown for different bias voltage V. Here $\Gamma_{0}= v/2 = 0.1$ meV, T = 1 K, B = 1 T, $p_{L}=p_{L}=0$ and $\varepsilon_{0}=0$ meV.}
	\label{voltage}
\end{figure}

\subsection{Non-magnetic leads}
Next, we study the time-dependent solution of the charge and spin currents and the evolution of the local spin moment for non-magnetic leads, $p_{\chi}=0$. Here, we use $\Gamma_{0}=v= 0.1$ meV, T = 1 K, B = 1 T, $p_{L}=p_{L}=0$, $\varepsilon_{0}=0$ meV and V = 2 mV. Our computed charge current $I^C$ is shown in Fig. \ref{voltage} for step-like voltage biases $V$ with different amplitudes, turned on at time $t=0$. The contour in Fig. \ref{voltage} (a) shows the time-evolution of $I^C(t)$ as function of $V$, while the plots in Fig. \ref{voltage} (b) correspond to the traces indicated in panel (a). The current acts as a response function to the step-like voltage bias according to a well known and expected scheme and eventually reaches the stationary regime.
A direct influence of amplitude of the voltage bias is the increasing the frequency of the current oscillations as the voltage bias grows as well as increasing decay time.
For non-magnetic leads the influence of the local magnetic moment $\bfS(t)$ on the current is negligible. Hence, the essential time-dependent properties of the charge current are captured by results provided in Refs. \cite{PhysRevB.48.8487, PhysRevB.50.5528}.

\begin{figure*}
	\centering
	\includegraphics[width=\textwidth]{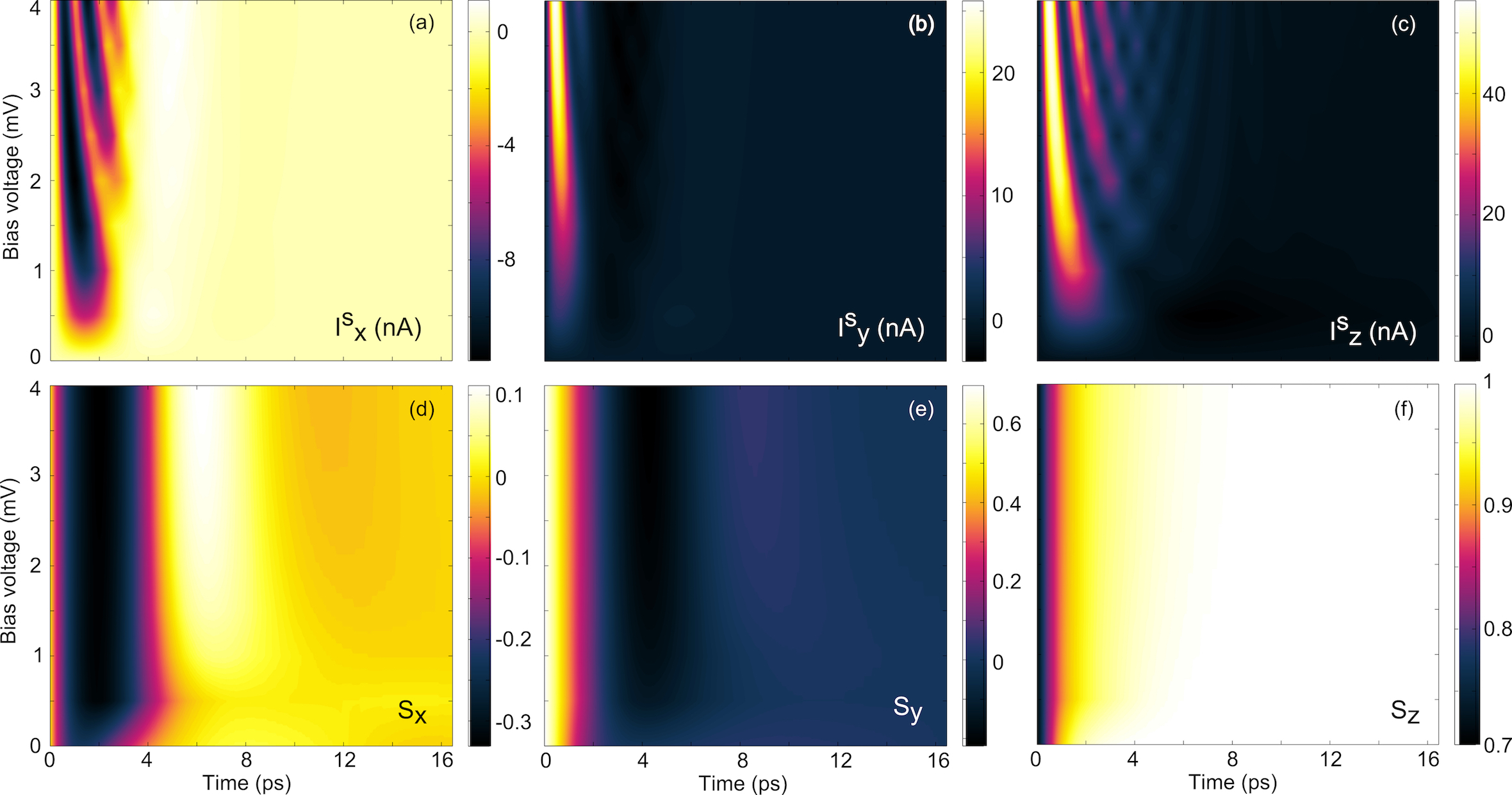}	
	\caption{Spin currents $\mathbb{I}_{S}(t)$ as a function of time for different step-like bias voltage (a) $I^{S}_{Lx}$, (b) $I^{S}_{Ly}$ and (c) $I^{S}_{Lz}$. Normalized local magnetic moment $\textbf{S}(t)$ as a function of time for different step-like bias voltage (d) $S_{x}$, (e) $S_{y}$ and (f) $S_{z}$. Here the same values as in Fig. \ref{voltage} is used.}
	\label{voltagespin}
\end{figure*}



In the case of non-magnetic leads, both $\bfGamma_{1} = 0$ and $\mathbf{g}_{1} = 0$, thus the only contribution to the spin-dependent dressed GF comes from $-vg_0\av{\bfS}g_0$. With no applied gate voltage this will lead to zero spin currents in the stationary limit, cf. Fig. \ref{timeindependent} (b). This can be seen in Fig. \ref{voltagespin} (a) - (c) where we show the spin currents in the x-, y-, and z-direction for different bias voltages. 
The local spin moment, initially at a polar angle $\pi/4$, will align in the z-direction due to the damping effect of the charge background, see Fig. \ref{voltagespin} (d)-(e). The process will be slower for a finite bias voltage than for zero voltage as there are anisotropic effects in the system. The back-action via the current in the junction causes this slower dynamics as it counteracts the motion in Fig. \ref{spinparameters} (notice that we have zero gate voltage). Here, the induced internal magnetic field and DM interaction causes the spin to align in the positive z-direction, whereas the Heisenberg and Ising interactions cause it to flip in the negative z-direction. 
Similar to this, the transient effects  in the spin currents in Fig. \ref{voltagespin} (a) - (c) are due to both the presence of the local spin moment and the current through the system, as it depends on $\bfG_1=-vg_0\av{\bfS}g_0$. As the dynamics of the local spin moment has a small dependence of the voltage bias, the time- and voltage-dependent changes of the spin current are thus mainly due to the electron flow in the system, given by $g_0$.

\begin{figure*}
	\centering
	\includegraphics[width=\textwidth]{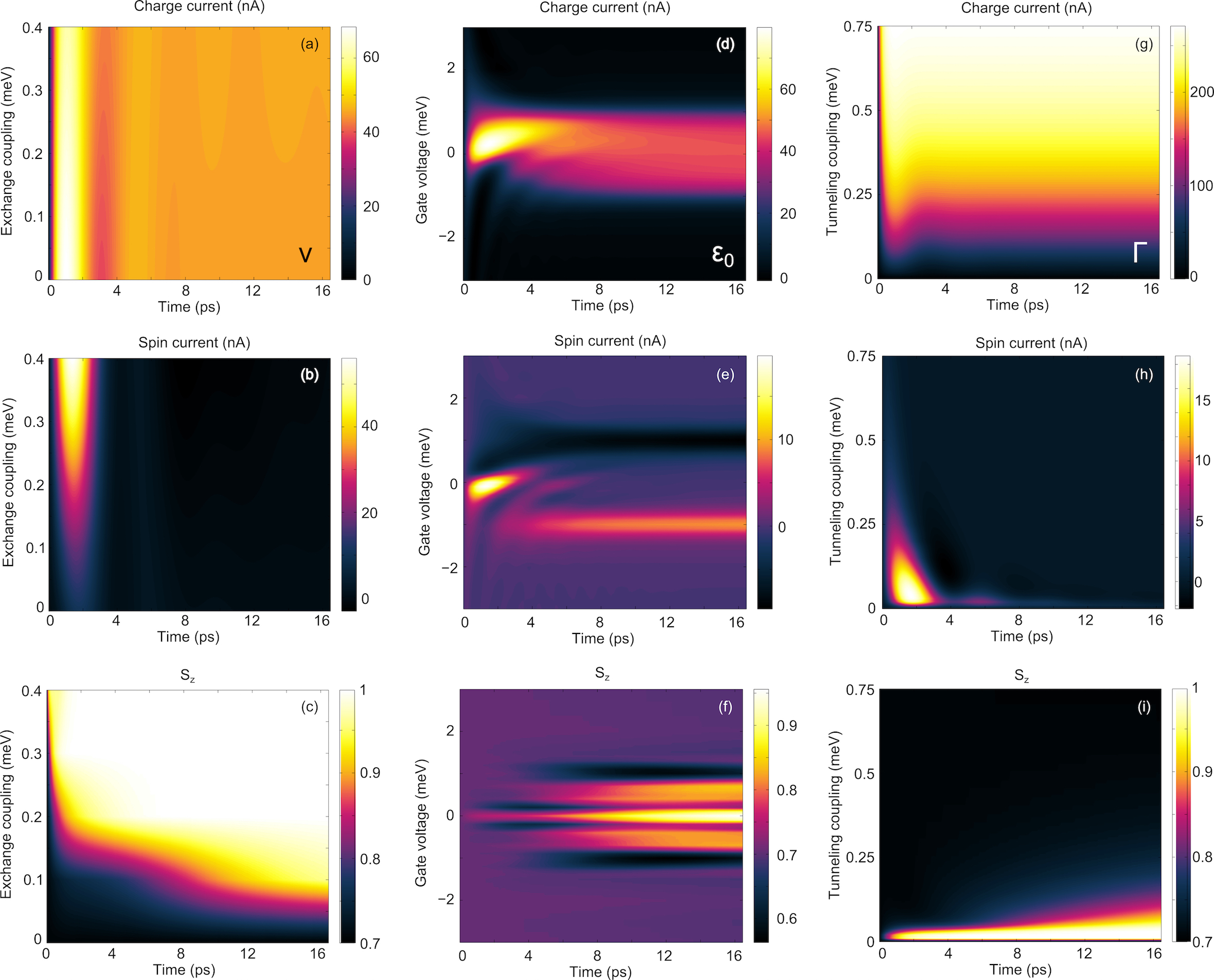}	
	\caption{Evolution of the currents and local magnetic moment when adjusting different parameters. For different exchange coupling, $v$, the figures show (a) the charge current  ${\rm I^{C}}$, (b) the spin current $I^{S}_{z}$ and (c) the local magnetic moment in the z-direction $S_{z}$. For different gate voltage, $\varepsilon_{0}$, the figures show (d) the charge current ${\rm I^{C}}$, (e) the spin current $I^{S}_{z}$ and (f) the local magnetic moment in the z-direction $S_{z}$. For different tunneling coupling, $\Gamma_{0}$, the figures show (g) the charge current ${\rm I^{C}}$, (h) the spin current $I^{S}_{z}$ and (i) the local magnetic moment in the z-direction $S_{z}$. Here the same values as in Fig. \ref{timedependentparameters} is used.}
	\label{totalfigur}
\end{figure*}

If there is no exchange coupling between the QD and the local spin moment, i.e. $v$ = 0, the spin moment would just continue to rotate in the magnetic field and there would not be any induced spin currents in the system. This can be seen in Fig. \ref{totalfigur} (b)-(c) for a finite charge current through the system in Fig. \ref{totalfigur} (a). The interaction between the current through the QD and the local spin moment increases as we turn on the exchange coupling, thus changing the direction of the spin. The scaling behavior depends on the interaction between the QD and the local spin moment. The internal magnetic field scales linearly with the exchange coupling, $\mathbf{\epsilon j} \propto v$, and the current mediated interaction scales quadratically, $\mathbb{J} \propto v^2$. In turn, these equations depends on the back-action from the spin moment through the GF, defined as $\bfG_1=-vg_0\av{\bfS}g_0$. The scaling of the exchange coupling is thus $v^{4}$ and in Fig. \ref{totalfigur} (c) we can observe a 16 times faster process for every doubling of the exchange coupling.
The spin current Fig. \ref{totalfigur} (b) scales linearly with exchange coupling, due to $\bfG_1=-vg_0\av{\bfS}g_0$, and the charge current Fig. \ref{totalfigur} (a) is independent of the exchange coupling as it only depends on $g_0$ for non-magnetic leads.

The QD electron level $\epsilon_{0}$ is adjusted by the gate voltage V. In the stationary limit, the charge current peaks when the QD electron level lies between the chemical potential of the leads, and it quickly diminishes for higher and lower gate voltage. Because of the effective Zeeman split in the QD there will be finite spin currents that are strongly peaked at $\mu_\chi=\dote{0}$. This can also be observed in the long time limit for different gate voltage, shown in Fig. \ref{totalfigur} (d)-(e). Due to the asymmetry of the time-dependent solution of the bare GF, where $g_0(t,t')$ is not the same as $g_0(t',t)$ around the onset of bias voltage, the currents become asymmetric for small time scales depending on the sign of the gate voltage. As this is a short-term effect, the asymmetry vanishes as it reaches steady state.
The evolution of the local spin moment, shown in Fig. \ref{spinparameters}, depends on the gate voltage as the strength of the interaction strongly depends on the gate voltage, see Fig. \ref{timedependentparameters} and Fig. \ref{spinparameters}. For zero gate voltage, the local spin moment simply aligns with the leads, as in Fig. \ref{voltagespin} (f). When a finite gate voltage is applied, the QD level is not symmetrically in between the leads, which gives both a spin-polarized current due to the Zeeman split and in turn changes the action on the spin moment. This causes the spin moment to reach different solutions in the steady state depending on both the isotropic and anisotropic interactions shown in Fig. \ref{spinparameters}.

Recalling that the charge current for non-magnetic leads is given in the stationary limit by Eq. \ref{charge current stationary} we notice that it scales with the tunneling coupling $\Gamma_{0}$ as an Lorentzian. This can be seen in the steady state limit in Fig. \ref{totalfigur} (g) where the charge current is plotted against different tunneling coupling. The dependence of $\Gamma_{0}$ in both the spin currents and the interaction strength is as in Eqs. \ref{eq-HC1} -- \ref{eq-JG1}, where it is of the form $\Gamma_{0}/(1+\Gamma_{0}^{2})^{2}$, $\Gamma_{0}/(1+\Gamma_{0}^{4})^{2}$ and $\Gamma_{0}^{3}/(1+\Gamma_{0}^{4})^{2}$. This will give a high contribution for a narrow range and can be seen in Fig. \ref{totalfigur} (h)-(i), where the spin current and evolution of the local spin moment is plotted against different tunneling coupling.

\begin{figure*}
	\centering
	\includegraphics[width=\textwidth]{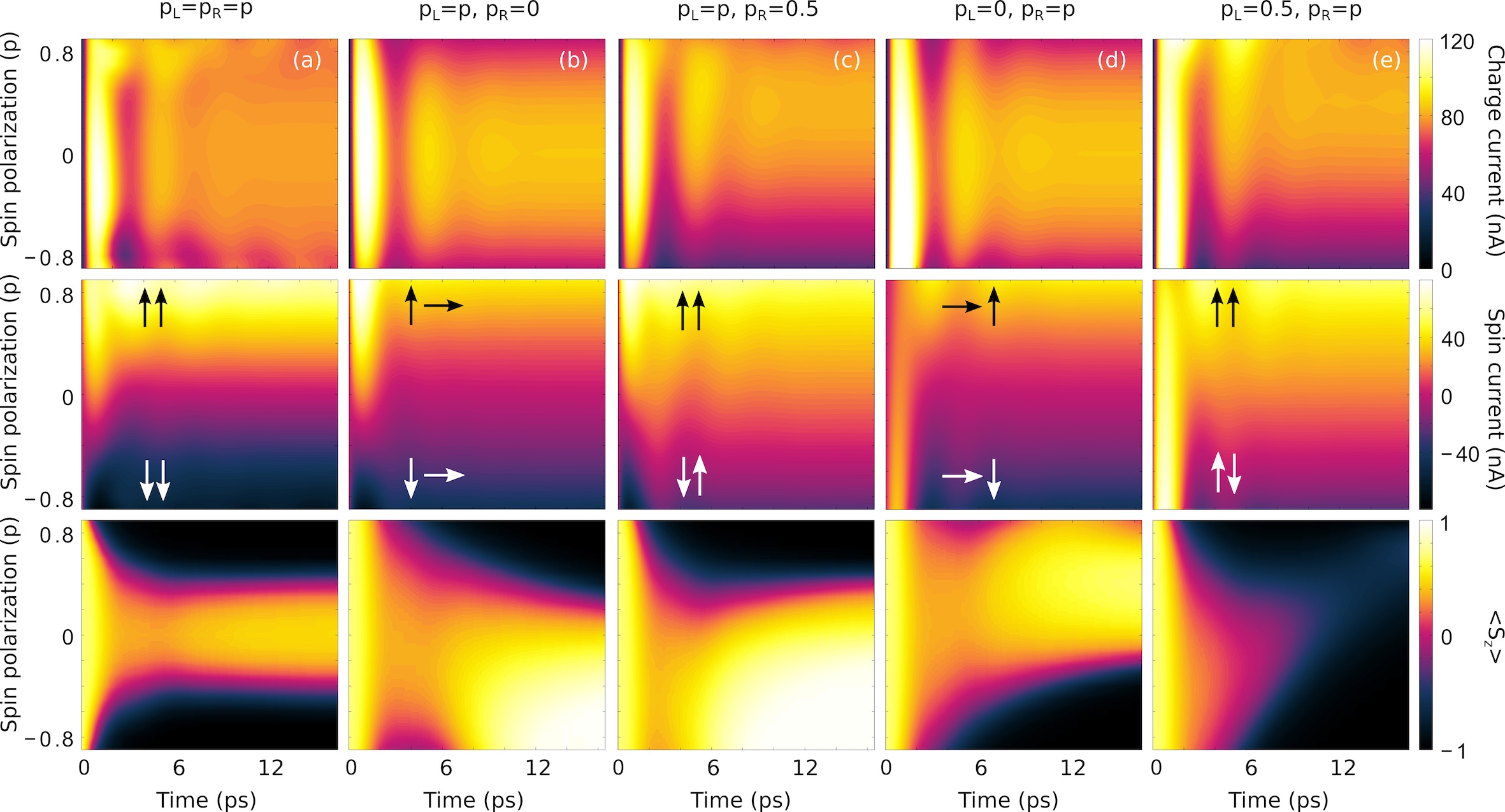}	
	\caption{Evolution of the currents and local magnetic moment for different polarization p of the leads indicated by the arrows in the second row, where the left arrow indicates the polarization of the left lead and the right arrow indicates the right lead. The first row shows the charge current ${\rm I^{C}}$, the second row the spin currents $I^{S}_{z}$ and the third row shows the local magnetic moment in the z-direction $S_{z}$. In column (a) the magnetization is changed for both leads, $p_{L}=p_{R}=p$. In column (b) the right lead is non-magnetic and the left lead is changed, $p_{R}=0$ and $p_{L}=p$. In column (c) the right lead is ferromagnetic and the left lead is changed, $p_{R}=0.5$ and $p_{L}=p$. In column (d) the left lead is non-magnetic and the right lead is changed, $p_{R}=p$ and $p_{L}=0$. In column (e) the left lead is ferromagnetic and the right lead is changed, $p_{R}=p$ and $p_{L}=0.5$.  Here $\Gamma_{0}= 0.1$ meV, $v= 0.25$ meV, V = 2 mV, T = 1 $\mu K$, B = 1 $\mu T$, $p_{L}=p_{L}=0$ and $\epsilon_{0}=0$ meV.}
	\label{polarization}
\end{figure*}

\subsection{Ferromagnetic leads}
We now study the system for different magnetization of the leads. In Fig. \ref{polarization} the different columns show different spin polarization of the leads. To enhance the effects, the simulation was run with T = 1 $\mu$K and B = 1 $\mu$T. In Fig. \ref{polarization} column (a) the magnetization is changed for both leads, $p_{L}=p_{R}=p$, in a ferromagnetic alignment. As can be seen in the stationary limit, the spin current is then net positive or net negative due to the spin polarization of the leads, as expected. The charge current decreases some for strongly polarized leads as one of the spin species diminishes, thus only tunneling through either the spin-up or -down channel (see Fig. \ref{simplesystem} (b) for illustration). 
If the drain lead is kept non-magnetic, i.e. $p_{L}=p, p_{R}=0$,  this behavior becomes clearer; see Fig. \ref{polarization} column (b). Here, a majority of spin-up or spin-down electrons enters the QD while both have equal probability to exit. The spin current through the junction becomes less polarized because only one of the leads is ferromagnetic, while the other is non-magnetic. The changes in charge current are large, because of the non-collinear arrangement of the leads. Similar changes in charge and spin currents appear when the source is kept non-magnetic, i.e. $p_{L}=0, p_{R}=p$, see Fig. \ref{polarization} column (d).

When the source or drain lead is kept ferromagnetic, i.e., $p_{\chi}=0.5$, the charge and spin current behavior changes slightly; see Fig. \ref{polarization} column (c) and (e). We get the highest charge current when the leads are in a ferromagnetic configuration, while it is the lowest in an anti-ferromagnetic configuration. The same happens for the spin current, where it is the highest in a ferromagnetic configuration and it goes to zero/negative when the leads are in an anti-ferromagnetic configuration. This is well expected and agrees with a simple model of a QD between magnetic leads.

\begin{figure}
	\includegraphics[width=\columnwidth]{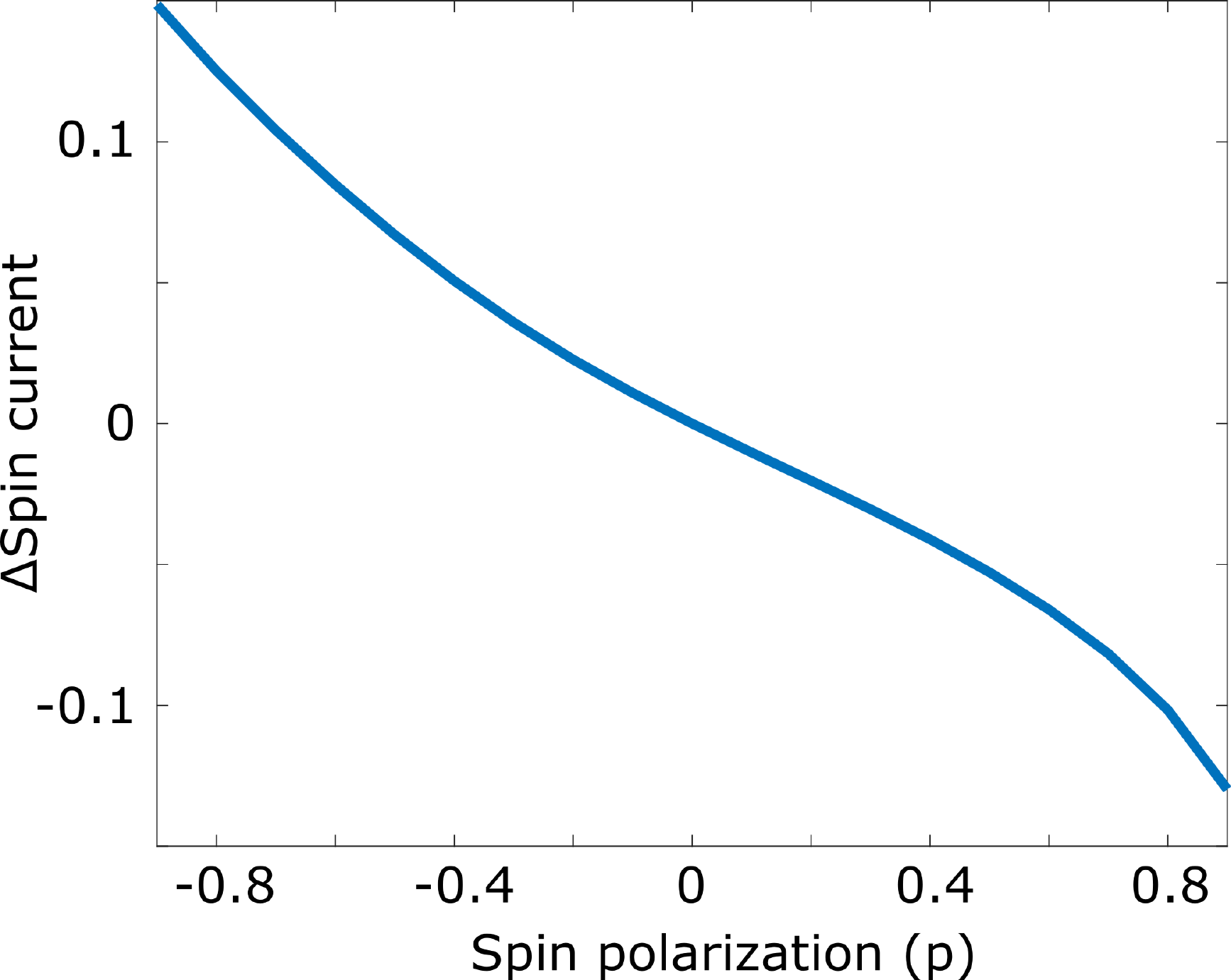}	
	\caption{Difference in spin current normalized by charge current in the stationary limit, $[I^{S}_{z}(p_{L}=0,p_{R}=p)-I^{S}_{z}(p_{L}=p,p_{R}=0)]/I^{C}$,  depending on direction of current with one lead non-magnetic and the other lead with different spin polarization. This plot corresponds to the difference between columns (b) and (d) in the second row of Fig. \ref{polarization}. The difference is due to the different directions of the local spin moment.}
	\label{diffspincurrent}
\end{figure}

We notice that there is a slight difference between the spin currents in Fig. \ref{polarization} column (b) and (d) and the same for column (c) and (e). This is partly due to the asymmetry of the source and drain leads, which causes an effect on small time scales when the bias voltage is turned on, where a spin-up injection causes a peak in the spin current, and a spin-down injection causes a bottom. It is also due to the influence of the local spin moment in the junction which reaches different stationary solutions (see the bottom row of Fig. \ref{polarization}), as the dressed GF depends on the connection with the spin moment, as is given in Eq. \ref{eq:Dressed Greens function}. The difference is shown in Fig. \ref{diffspincurrent} for the case in which one lead is kept non-magnetic and the spin polarization of the other lead is shifted. The plot shows the difference $[I^{S}_{z}(p_{L}=0,p_{R}=p)-I^{S}_{z}(p_{L}=p,p_{R}=0)]/I^{C}$. Hence, because of the given polarity of the voltage bias, the system is not invariant with respect to inversion symmetry, and the configurations $p_{L}=0, p_{R}=p$ and $p_{L}=p, p_{R}=0$ are not equivalent under finite voltage bias.

\begin{figure}
	\includegraphics[width=\columnwidth]{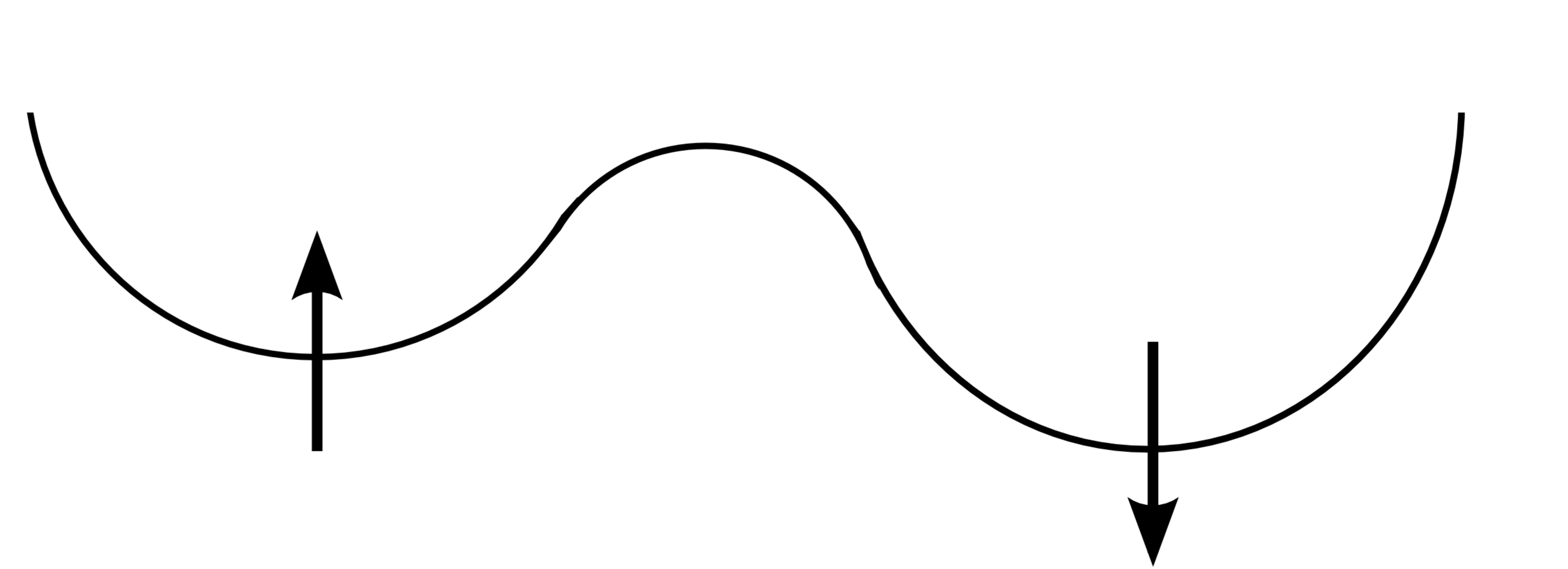}	
	\caption{Double well potential illustrating the possible solutions for the local spin moment depending on starting position. As can be seen in Fig. \ref{polarization} the spin moment starts up and then tries to reach a down solution. Depending on strength of the anisotropies in the system it either passes the energy barrier or not.}
	\label{doublewell}
\end{figure}

For ferromagnetic leads, the interaction between the QD and the spin moment changes. This causes the local spin moment to change direction for different spin-polarized configurations of the leads. The evolution of the spin moment for different configurations is shown in the bottom row in Fig. \ref{polarization}. Due to the anisotropic interactions $\mathbb{I}$ and \textbf{D}, the local spin moment anti-aligns with the source lead while it aligns with the drain lead; see  Figs. \ref{polarization}(b) and \ref{polarization}(c). The effect is stronger from the source lead than the drain lead, as the source determines the electron flow into the system.
In the small-time scale, right after the gate voltage is turned on, we can observe a dip in the evolution of the spin moment before its reaches its stationary solution. This we can describe energetically as a double well potential, where the spin moment needs enough energy in order to pass a barrier and reach a spin-down solution. If it does not have enough energy to pass the barrier, it stays in a spin-up solution; see Fig. \ref{doublewell}. In the present case, the spin is considered classical and the transition will be continuous, while a quantum spin would tunnel through the potential barrier for high enough energies. As we see in Figs. \ref{polarization}(b) and  \ref{polarization}(c), this configuration favours a spin up solution, while when the current is driven in the opposite direction, shown in  Figs. \ref{polarization}(d) and \ref{polarization}(e), it favors a spin-down solution. This is illustrated by the different depths of the potential wells in Fig.  \ref{doublewell}. 
Due to the anisotropies in the system, the spin moment ends up differently depending on the anti-ferromagnetic configurations of the leads. The anti-ferromagnetic configuration is the opposite in the bottom of the figures in (c) and (e) in the bottom row in Fig. \ref{polarization}, which will cause different stationary solutions of the local spin moment. Thus, depending on the direction of the current through the dot one can use the effective spin torque in order to control the spin, which is in agreement with previous studies \cite{Misiorny:2013aa}.

\section{Summary and conclusion}
\label{sec-summary}
In summary, we have studied the time evolution of a local magnetic moment in a tunnel junction. We have shown that one can control and read-out the local magnetic moment using gate voltage and using magnetic leads in an anti-ferromagnetic set-up. This is in agreement with previous works\cite{PhysRevB.73.235304,Misiorny:2013aa,PhysRevB.75.134425} and is a promising feature in order to perform electrical control and read-out of magnetic molecules.

We have shown that non-trivial exchange interaction appears in the time-dependent domain, especially for small time scales. Anisotropic effects occur due to time-dependency which will effect the direction of the magnetic moment. A large effective magnetic field is a significant effect that occurs for small time scales and adjusts the evolution of the local magnetic moment, an effect not usually considered as it vanishes for the stationary solution. Considering time-dependent exchange interaction is thus important in small time-scale calculations and shows the potential for a deeper understanding of the exchange interaction. This leads to further questions on how important the time-dependency is for large scale spin-dynamics calculations, something suitable for further investigation.

This work and previous studies have found that there is rich physics within this framework and that it is important on the quantum scale to take time-dependent non-equilibrium effects into consideration when analyzing time-dependent phenomena. Anisotropic exchange interactions play an important role when studying time-dependent phenomena. With recent experimental advances, we believe it to be possible to verify our findings with state-of-the-art experiments.



\acknowledgments
We thank J. -X. Zhu for inspiring discussions in the initial stages of this work. We also thank K. Bj\"ornson, T. L\"othman and J. D. Vasquez Jaramillo for fruitful discussions and for comments on our results. Financial support from Vetenskapsr\aa det is acknowledged. This work is part of the CINT User Proposal \# U2015A0056.

\appendix
\section{Effective action and stochastic fields}
\label{app-rv}
We use the closed time-path Green function (CTPGF) formalism \cite{CHOU19851} and thus calculate the partition function (with closed path contour ordering operator $T_C$),
\begin{align}
{\cal Z}[\bfS_n(t)]=&
	\tr T_C\exp{[i{\cal S}]},
\\
{\cal S}=&
	{\cal S}_\text{\tiny{WZWN}}
	+
	\oint_C
		\Hamil
	dt,
\end{align}
Here, ${\cal S}_\text{\tiny{WZWN}}=\int\bfS^q(t)\cdot[\bfS^c(t)\times\dt\bfS^c(t)]dt/ |{\bf S}|^2$ is the Wess-Zumino-Witten-Novikov (WZWN) term describing the Berry phase accumulated by the local spins. The trace is taken over the conduction electrons in the two leads in order to provide an effective spin action, which in the present situation represents the interaction of the magnetic spins with a non-equilibrium environment.

As the modell is defined on the Keldysh contour, which is necessary for general non-equilibrium conditions, we have to keep track of whether a spin operator is defined for times on the upper (lower) part of the contour. We do this by assigning the superscripts $u$ ($l$). Then, it is convenient to define the new spin operators $\bfS^c=(\bfS^u+\bfS^l)/2$ and $\bfS^q=\bfS^u-\bfS^l$ and following the procedure in Ref. \cite{franssonNJP2008}, we find that the effective action can be written as
\begin{align}
{\cal S}=&
	{\cal S}_\text{\tiny{WZWN}}
	+g\mu_B\int\bfB(t)\cdot\bfS^q(t)dt
	+\frac{1}{e}\int\boldsymbol{\epsilon}\bfj(t)\cdot\bfS^q(t)dt
\nonumber\\&
	+\frac{1}{e}\int\bfS^q(t)\cdot\mathbb{J}(t,t')\cdot\bfS^c(t')dtdt'
\nonumber\\&
	+\frac{1}{e}\int\bfS^q(t)\cdot\mathbb{J}^K(t,t')\cdot\bfS^q(t')dtdt'.
\end{align}
Here, the fields $\boldsymbol{\epsilon}\bfj(t)$ and $\mathbb{J}(t,t')$ are given in the main text, whereas $\mathbb{J}^K(t,t')=iev^2\av{\anticom{\bfs(t)}{\bfs(t')}}$ defines the electronically mediated interactions between the spin operators $\bfS^q(t)$ and $\bfS^q(t')$. This coupling between $\bfS^q(t)$ and $\bfS^q(t')$ provides a contribution to the model which is of different nature than the one between $\bfS^q(t)$ and $\bfS^c(t')$.

The operators $\bfS^c$ and $\bfS^q$ can be considered as slow and fast variables and the resulting equation of motion presented in this paper concerns the dynamics of the slow variable. The equation of motion is obtained from the saddle-point solution of the functional derivative of ${\cal S}$ with respect to $\bfS^q$, that is, by differentiating out the rapid dynamics from the discussion. Concerning the first four terms in the action, this leaves a model that is linear in the slow variable $\bfS^c$, and by finally cross-multiplying from the left by $\bfS^c$ under the assumption that $\partial_t|\bfS^c|^2=0$, we retain the equation of motion given in Eq. \ref{spinequationofmotion}.

Regarding the last contribution to the action, however, the functional differentiation results in a term that is linear in $\bfS^q$. In order to handle this complication there are, at least, two routes to solving the problem. First, one can also consider an equation of motion for the rapid dynamics, as well, which results in a coupled system of equations for the slow and fast variables. The second route would be to perform a Hubbard-Stratonovich transformation of the term in the action that is bilinear in $\bfS^q$. By this procedure, the problem is linearized at the price of introducing a stochastic field, represented by the Gaussian random variables $\xi(t)$. These random fields provides a quantum fluctuation description of the spin correlations.

The second route leads to the fact that Eq. (\ref{spinequationofmotion}) also contains a contribution of the form $\gamma\bfS(t)\times\xi(t)$, which can be interpreted as a random magnetic field acting on the spin. It can be shown that the random variable $\xi$ is defined by the electronic correlations through $(g\mu_B)^2\av{\xi(t)\xi(t')}=-i2\mathbb{J}^K(t,t')/e$ \cite{franssonNJP2008}. In the wide band limit for the electronic states, it is easy to show that the spin-spin correlation function $\mathbb{J}^K(t,t')\propto\delta(t-t')$, which shows that the stochastic field is of Gaussian white noise character in our set-up.

\section{Derivation of the Landau-Lifshitz-Gilbert equation}
\label{app-LLGderivation}

The Landau-Lifshitz-Gilbert (LLG) equation is usually defined as
\begin{align}
\mathbf{\dot{S}}=\mathbf{S}\times(-\gamma\mathbf{B}+\mathbf{\hat{G}}\mathbf{\dot{S}})
\end{align}
where \textbf{B} is the effective magnetic field acting on the spin, and $\mathbf{\hat{G}}$ is the Gilbert damping. Starting from the spin equation of motion used in this work, defined in Eq. \ref{spinequationofmotion}, and restriction to the adiabatic limit, we can retain the LLG equation \cite{PhysRevLett.108.057204}. In this limit one can assume that the spin is slowly varying with time, which allows to Taylor expand according to $t'$, i.e. $\mathbf{S}(t') = \mathbf{S}(t) - (t-t')\mathbf{\dot{S}}(t) + \mathcal{O}(\mathbf{\ddot{S}}(t))$. From this we obtain
\begin{align}
\int \mathbb{J}(t,t')\cdot\mathbf{S}(t')dt' \approx&
		\int \mathbb{J}(t,t')dt'\mathbf{S}(t)
\nonumber\\&
		-
		\int \mathbb{J}(t,t')(t-t')dt'\mathbf{\dot{S}}(t)
	.
\end{align}
Here, the first term corresponds to the internal magnetic field due to the spin background whereas the second term is related to the Gilbert damping. The equation of motion then simplifies to 
\begin{align}
\dot{\mathbf{S}}(t) = \mathbf{S}(t)\times\left( -g\mu_{B}\mathbf{B}^\text{eff}(t)+\mathbf{\hat{G}}\mathbf{\dot{S}}(t)\right)
\end{align}
where the first term represents the total effective magnetic field
\begin{align}
\mathbf{B}^\text{eff}(t) = \textbf{B}
+
\frac{1}{eg\mu_{B}}\int\boldsymbol{\epsilon}\textbf{j}(t,t')dt' + 
\frac{1}{eg\mu_{B}}\int \mathbb{J}(t,t')dt'\mathbf{S}(t)
,
\end{align}
whereas the Gilbert damping is given by
\begin{equation}
\mathbf{\hat{G}} = -\frac{1}{e}\int \mathbb{J}(t,t')(t-t')dt'.
\end{equation}

\section{Lesser/greater dressed quantum Green function}
\label{app-NEGF}
The lesser/greater forms of the correction to the dressed GF becomes
\begin{align}
\delta\bfG^{</>}(t,t')=&
	-
	v\int
	\Bigl(
		\bfg^r(t,\tau)\av{\bfS(\tau)}\cdot\bfsigma\bfg^{</>}(\tau,t')
\nonumber\\&
		+
		\bfg^{</>}(t,\tau)\av{\bfS(\tau)}\cdot\bfsigma\bfg^a(\tau,t')
	\Bigr)
	d\tau
	,
\label{eq:Greens function}
\end{align}
and decomposing into the charge and magnetic components we have
\begin{widetext}
\begin{subequations}
\begin{align}
\delta G_{0}^{</>}(t,t')=&
	-v\int\left(g_{0}^{r}(t,\tau)\left\langle \mathbf{S}(\tau)\right\rangle \mathbf{\cdot g_{1}^{</>}}(\tau,t')
		s+g_{0}^{</>}(t,\tau)\left\langle \mathbf{S}(\tau)\right\rangle \mathbf{\cdot}\mathbf{g}_{1}^{a}(\tau,t')
		\mathbf{+g}_{1}^{r}(t,\tau)\cdot\left\langle \mathbf{S}(\tau)\right\rangle g_{0}^{</>}(\tau,t')\right.
\nonumber\\&
		\left. +\mathbf{g}_{1}^{</>}(t,\tau)\cdot\left\langle \mathbf{S}(\tau)\right\rangle g_{0}^{a}(\tau,t')
		+i\left[\mathbf{g}_{1}^{r}(t,\tau)\times\left\langle \mathbf{S}(\tau)\right\rangle \right]\mathbf{\cdot}\mathbf{g}_{1}^{</>}(\tau,t')
		+i\left[\mathbf{g}_{1}^{</>}(t,\tau)\times\left\langle \mathbf{S}(\tau)\right\rangle \right]\mathbf{\cdot}\mathbf{g}_{1}^{a}(\tau,t')\right)d\tau,
\\
\mathbf{G}_{1}^{</>}(t,t')=&
	-v\int\left(g_{0}^{r}(t,\tau)\left\langle \mathbf{S}(\tau)\right\rangle g_{0}^{</>}(\tau,t')
		+g_{0}^{</>}(t,\tau)\left\langle \mathbf{S}(\tau)\right\rangle g_{0}^{a}(\tau,t')
		+i\left[\mathbf{g}_{1}^{r}(t,\tau)\times\left\langle \mathbf{S}(\tau)\right\rangle \right]g_{0}^{</>}(\tau,t')\right.
\nonumber \\&
		\left. +i\left[\mathbf{g}_{1}^{</>}(t,\tau)\times\left\langle \mathbf{S}(\tau)\right\rangle \right]g_{0}^{a}(\tau,t')
		+ig_{0}^{r}(t,\tau)\left[\left\langle \mathbf{S}(\tau)\right\rangle \times\mathbf{g}_{1}^{</>}(\tau,t')\right]
		+ig_{0}^{</>}(t,\tau)\left[\left\langle \mathbf{S}(\tau)\right\rangle \times\mathbf{g}_{1}^{a}(\tau,t')\right]\right.
\nonumber \\&
		+\left.i\left[\mathbf{g}_{1}^{r}(t,\tau)\times\left\langle \mathbf{S}(\tau)\right\rangle \right]\mathbf{\times}\mathbf{g}_{1}^{</>}(\tau,t')
		+i\left[\mathbf{g}_{1}^{</>}(t,\tau)\times\left\langle \mathbf{S}(\tau)\right\rangle \right]\mathbf{\times}\mathbf{g}_{1}^{a}(\tau,t')\right)d\tau.
\end{align}
\end{subequations}
\end{widetext}


\begin{thebibliography}{62}
	\expandafter\ifx\csname natexlab\endcsname\relax\def\natexlab#1{#1}\fi
	\expandafter\ifx\csname bibnamefont\endcsname\relax
	\def\bibnamefont#1{#1}\fi
	\expandafter\ifx\csname bibfnamefont\endcsname\relax
	\def\bibfnamefont#1{#1}\fi
	\expandafter\ifx\csname citenamefont\endcsname\relax
	\def\citenamefont#1{#1}\fi
	\expandafter\ifx\csname url\endcsname\relax
	\def\url#1{\texttt{#1}}\fi
	\expandafter\ifx\csname urlprefix\endcsname\relax\def\urlprefix{URL }\fi
	\providecommand{\bibinfo}[2]{#2}
	\providecommand{\eprint}[2][]{\url{#2}}
	
	\bibitem[{\citenamefont{Hauptmann et~al.}(2008)\citenamefont{Hauptmann, Paaske,
			and Lindelof}}]{Hauptmann:2008aa}
	\bibinfo{author}{\bibfnamefont{J.~R.} \bibnamefont{Hauptmann}},
	\bibinfo{author}{\bibfnamefont{J.}~\bibnamefont{Paaske}}, \bibnamefont{and}
	\bibinfo{author}{\bibfnamefont{P.~E.} \bibnamefont{Lindelof}},
	\bibinfo{journal}{Nat Phys} \textbf{\bibinfo{volume}{4}},
	\bibinfo{pages}{373} (\bibinfo{year}{2008}).
	
	\bibitem[{\citenamefont{Loth et~al.}(2010)\citenamefont{Loth, von Bergmann,
			Ternes, Otte, Lutz, and Heinrich}}]{Loth:2010aa}
	\bibinfo{author}{\bibfnamefont{S.}~\bibnamefont{Loth}},
	\bibinfo{author}{\bibfnamefont{K.}~\bibnamefont{von Bergmann}},
	\bibinfo{author}{\bibfnamefont{M.}~\bibnamefont{Ternes}},
	\bibinfo{author}{\bibfnamefont{A.~F.} \bibnamefont{Otte}},
	\bibinfo{author}{\bibfnamefont{C.~P.} \bibnamefont{Lutz}}, \bibnamefont{and}
	\bibinfo{author}{\bibfnamefont{A.~J.} \bibnamefont{Heinrich}},
	\bibinfo{journal}{Nat Phys} \textbf{\bibinfo{volume}{6}},
	\bibinfo{pages}{340} (\bibinfo{year}{2010}).
	
	\bibitem[{\citenamefont{Wagner et~al.}(2013)\citenamefont{Wagner, Kisslinger,
			Ballmann, Schramm, Chandrasekar, Bodenstein, Fuhr, Secker, Fink, Ruben
			et~al.}}]{Wagner:2013aa}
	\bibinfo{author}{\bibfnamefont{S.}~\bibnamefont{Wagner}},
	\bibinfo{author}{\bibfnamefont{F.}~\bibnamefont{Kisslinger}},
	\bibinfo{author}{\bibfnamefont{S.}~\bibnamefont{Ballmann}},
	\bibinfo{author}{\bibfnamefont{F.}~\bibnamefont{Schramm}},
	\bibinfo{author}{\bibfnamefont{R.}~\bibnamefont{Chandrasekar}},
	\bibinfo{author}{\bibfnamefont{T.}~\bibnamefont{Bodenstein}},
	\bibinfo{author}{\bibfnamefont{O.}~\bibnamefont{Fuhr}},
	\bibinfo{author}{\bibfnamefont{D.}~\bibnamefont{Secker}},
	\bibinfo{author}{\bibfnamefont{K.}~\bibnamefont{Fink}},
	\bibinfo{author}{\bibfnamefont{M.}~\bibnamefont{Ruben}},
	\bibnamefont{et~al.}, \bibinfo{journal}{Nat Nano}
	\textbf{\bibinfo{volume}{8}}, \bibinfo{pages}{575} (\bibinfo{year}{2013}).
	
	\bibitem[{\citenamefont{Hirjibehedin et~al.}(2006)\citenamefont{Hirjibehedin,
			Lutz, and Heinrich}}]{Hirjibehedin19052006}
	\bibinfo{author}{\bibfnamefont{C.~F.} \bibnamefont{Hirjibehedin}},
	\bibinfo{author}{\bibfnamefont{C.~P.} \bibnamefont{Lutz}}, \bibnamefont{and}
	\bibinfo{author}{\bibfnamefont{A.~J.} \bibnamefont{Heinrich}},
	\bibinfo{journal}{Science} \textbf{\bibinfo{volume}{312}},
	\bibinfo{pages}{1021} (\bibinfo{year}{2006}).
	
	\bibitem[{\citenamefont{Wahl et~al.}(2007)\citenamefont{Wahl, Simon,
			Diekh\"oner, Stepanyuk, Bruno, Schneider, and Kern}}]{PhysRevLett.98.056601}
	\bibinfo{author}{\bibfnamefont{P.}~\bibnamefont{Wahl}},
	\bibinfo{author}{\bibfnamefont{P.}~\bibnamefont{Simon}},
	\bibinfo{author}{\bibfnamefont{L.}~\bibnamefont{Diekh\"oner}},
	\bibinfo{author}{\bibfnamefont{V.~S.} \bibnamefont{Stepanyuk}},
	\bibinfo{author}{\bibfnamefont{P.}~\bibnamefont{Bruno}},
	\bibinfo{author}{\bibfnamefont{M.~A.} \bibnamefont{Schneider}},
	\bibnamefont{and} \bibinfo{author}{\bibfnamefont{K.}~\bibnamefont{Kern}},
	\bibinfo{journal}{Phys. Rev. Lett.} \textbf{\bibinfo{volume}{98}},
	\bibinfo{pages}{056601} (\bibinfo{year}{2007}).
	
	\bibitem[{\citenamefont{Meier et~al.}(2008)\citenamefont{Meier, Zhou, Wiebe,
			and Wiesendanger}}]{Meier04042008}
	\bibinfo{author}{\bibfnamefont{F.}~\bibnamefont{Meier}},
	\bibinfo{author}{\bibfnamefont{L.}~\bibnamefont{Zhou}},
	\bibinfo{author}{\bibfnamefont{J.}~\bibnamefont{Wiebe}}, \bibnamefont{and}
	\bibinfo{author}{\bibfnamefont{R.}~\bibnamefont{Wiesendanger}},
	\bibinfo{journal}{Science} \textbf{\bibinfo{volume}{320}},
	\bibinfo{pages}{82} (\bibinfo{year}{2008}).
	
	\bibitem[{\citenamefont{Balashov et~al.}(2009)\citenamefont{Balashov, Schuh,
			Tak\'acs, Ernst, Ostanin, Henk, Mertig, Bruno, Miyamachi, Suga
			et~al.}}]{PhysRevLett.102.257203}
	\bibinfo{author}{\bibfnamefont{T.}~\bibnamefont{Balashov}},
	\bibinfo{author}{\bibfnamefont{T.}~\bibnamefont{Schuh}},
	\bibinfo{author}{\bibfnamefont{A.~F.} \bibnamefont{Tak\'acs}},
	\bibinfo{author}{\bibfnamefont{A.}~\bibnamefont{Ernst}},
	\bibinfo{author}{\bibfnamefont{S.}~\bibnamefont{Ostanin}},
	\bibinfo{author}{\bibfnamefont{J.}~\bibnamefont{Henk}},
	\bibinfo{author}{\bibfnamefont{I.}~\bibnamefont{Mertig}},
	\bibinfo{author}{\bibfnamefont{P.}~\bibnamefont{Bruno}},
	\bibinfo{author}{\bibfnamefont{T.}~\bibnamefont{Miyamachi}},
	\bibinfo{author}{\bibfnamefont{S.}~\bibnamefont{Suga}}, \bibnamefont{et~al.},
	\bibinfo{journal}{Phys. Rev. Lett.} \textbf{\bibinfo{volume}{102}},
	\bibinfo{pages}{257203} (\bibinfo{year}{2009}).
	
	\bibitem[{\citenamefont{Voss et~al.}(2008)\citenamefont{Voss, Zander, Fonin,
			R\"udiger, Burgert, and Groth}}]{PhysRevB.78.155403}
	\bibinfo{author}{\bibfnamefont{S.}~\bibnamefont{Voss}},
	\bibinfo{author}{\bibfnamefont{O.}~\bibnamefont{Zander}},
	\bibinfo{author}{\bibfnamefont{M.}~\bibnamefont{Fonin}},
	\bibinfo{author}{\bibfnamefont{U.}~\bibnamefont{R\"udiger}},
	\bibinfo{author}{\bibfnamefont{M.}~\bibnamefont{Burgert}}, \bibnamefont{and}
	\bibinfo{author}{\bibfnamefont{U.}~\bibnamefont{Groth}},
	\bibinfo{journal}{Phys. Rev. B} \textbf{\bibinfo{volume}{78}},
	\bibinfo{pages}{155403} (\bibinfo{year}{2008}).
	
	\bibitem[{\citenamefont{Bairagi et~al.}(2015)\citenamefont{Bairagi, Bellec,
			Repain, Chacon, Girard, Garreau, Lagoute, Rousset, Breitwieser, Hu
			et~al.}}]{PhysRevLett.114.247203}
	\bibinfo{author}{\bibfnamefont{K.}~\bibnamefont{Bairagi}},
	\bibinfo{author}{\bibfnamefont{A.}~\bibnamefont{Bellec}},
	\bibinfo{author}{\bibfnamefont{V.}~\bibnamefont{Repain}},
	\bibinfo{author}{\bibfnamefont{C.}~\bibnamefont{Chacon}},
	\bibinfo{author}{\bibfnamefont{Y.}~\bibnamefont{Girard}},
	\bibinfo{author}{\bibfnamefont{Y.}~\bibnamefont{Garreau}},
	\bibinfo{author}{\bibfnamefont{J.}~\bibnamefont{Lagoute}},
	\bibinfo{author}{\bibfnamefont{S.}~\bibnamefont{Rousset}},
	\bibinfo{author}{\bibfnamefont{R.}~\bibnamefont{Breitwieser}},
	\bibinfo{author}{\bibfnamefont{Y.-C.} \bibnamefont{Hu}},
	\bibnamefont{et~al.}, \bibinfo{journal}{Phys. Rev. Lett.}
	\textbf{\bibinfo{volume}{114}}, \bibinfo{pages}{247203}
	(\bibinfo{year}{2015}).
	
	\bibitem[{\citenamefont{Zhou et~al.}(2010)\citenamefont{Zhou, Wiebe, Lounis,
			Vedmedenko, Meier, Bl\"ugel, Dederichs, and Wiesendanger}}]{Zhou:2010aa}
	\bibinfo{author}{\bibfnamefont{L.}~\bibnamefont{Zhou}},
	\bibinfo{author}{\bibfnamefont{J.}~\bibnamefont{Wiebe}},
	\bibinfo{author}{\bibfnamefont{S.}~\bibnamefont{Lounis}},
	\bibinfo{author}{\bibfnamefont{E.}~\bibnamefont{Vedmedenko}},
	\bibinfo{author}{\bibfnamefont{F.}~\bibnamefont{Meier}},
	\bibinfo{author}{\bibfnamefont{S.}~\bibnamefont{Bl\"ugel}},
	\bibinfo{author}{\bibfnamefont{P.~H.} \bibnamefont{Dederichs}},
	\bibnamefont{and}
	\bibinfo{author}{\bibfnamefont{R.}~\bibnamefont{Wiesendanger}},
	\bibinfo{journal}{Nat Phys} \textbf{\bibinfo{volume}{6}},
	\bibinfo{pages}{187} (\bibinfo{year}{2010}).
	
	\bibitem[{\citenamefont{Chen et~al.}(2008)\citenamefont{Chen, Fu, Ji, Zhang,
			Cheng, Ma, Zou, Duan, Jia, and Xue}}]{PhysRevLett.101.197208}
	\bibinfo{author}{\bibfnamefont{X.}~\bibnamefont{Chen}},
	\bibinfo{author}{\bibfnamefont{Y.-S.} \bibnamefont{Fu}},
	\bibinfo{author}{\bibfnamefont{S.-H.} \bibnamefont{Ji}},
	\bibinfo{author}{\bibfnamefont{T.}~\bibnamefont{Zhang}},
	\bibinfo{author}{\bibfnamefont{P.}~\bibnamefont{Cheng}},
	\bibinfo{author}{\bibfnamefont{X.-C.} \bibnamefont{Ma}},
	\bibinfo{author}{\bibfnamefont{X.-L.} \bibnamefont{Zou}},
	\bibinfo{author}{\bibfnamefont{W.-H.} \bibnamefont{Duan}},
	\bibinfo{author}{\bibfnamefont{J.-F.} \bibnamefont{Jia}}, \bibnamefont{and}
	\bibinfo{author}{\bibfnamefont{Q.-K.} \bibnamefont{Xue}},
	\bibinfo{journal}{Phys. Rev. Lett.} \textbf{\bibinfo{volume}{101}},
	\bibinfo{pages}{197208} (\bibinfo{year}{2008}).
	
	\bibitem[{\citenamefont{Otte et~al.}(2009)\citenamefont{Otte, Ternes, Loth,
			Lutz, Hirjibehedin, and Heinrich}}]{PhysRevLett.103.107203}
	\bibinfo{author}{\bibfnamefont{A.~F.} \bibnamefont{Otte}},
	\bibinfo{author}{\bibfnamefont{M.}~\bibnamefont{Ternes}},
	\bibinfo{author}{\bibfnamefont{S.}~\bibnamefont{Loth}},
	\bibinfo{author}{\bibfnamefont{C.~P.} \bibnamefont{Lutz}},
	\bibinfo{author}{\bibfnamefont{C.~F.} \bibnamefont{Hirjibehedin}},
	\bibnamefont{and} \bibinfo{author}{\bibfnamefont{A.~J.}
		\bibnamefont{Heinrich}}, \bibinfo{journal}{Phys. Rev. Lett.}
	\textbf{\bibinfo{volume}{103}}, \bibinfo{pages}{107203}
	(\bibinfo{year}{2009}).
	
	\bibitem[{\citenamefont{Pruser et~al.}(2011)\citenamefont{Pruser, Wenderoth,
			Dargel, Weismann, Peters, Pruschke, and Ulbrich}}]{Pruser:2011aa}
	\bibinfo{author}{\bibfnamefont{H.}~\bibnamefont{Pruser}},
	\bibinfo{author}{\bibfnamefont{M.}~\bibnamefont{Wenderoth}},
	\bibinfo{author}{\bibfnamefont{P.~E.} \bibnamefont{Dargel}},
	\bibinfo{author}{\bibfnamefont{A.}~\bibnamefont{Weismann}},
	\bibinfo{author}{\bibfnamefont{R.}~\bibnamefont{Peters}},
	\bibinfo{author}{\bibfnamefont{T.}~\bibnamefont{Pruschke}}, \bibnamefont{and}
	\bibinfo{author}{\bibfnamefont{R.~G.} \bibnamefont{Ulbrich}},
	\bibinfo{journal}{Nat Phys} \textbf{\bibinfo{volume}{7}},
	\bibinfo{pages}{203} (\bibinfo{year}{2011}).
	
	\bibitem[{\citenamefont{Khajetoorians et~al.}(2011)\citenamefont{Khajetoorians,
			Wiebe, Chilian, and Wiesendanger}}]{Khajetoorians27052011}
	\bibinfo{author}{\bibfnamefont{A.~A.} \bibnamefont{Khajetoorians}},
	\bibinfo{author}{\bibfnamefont{J.}~\bibnamefont{Wiebe}},
	\bibinfo{author}{\bibfnamefont{B.}~\bibnamefont{Chilian}}, \bibnamefont{and}
	\bibinfo{author}{\bibfnamefont{R.}~\bibnamefont{Wiesendanger}},
	\bibinfo{journal}{Science} \textbf{\bibinfo{volume}{332}},
	\bibinfo{pages}{1062} (\bibinfo{year}{2011}).
	
	\bibitem[{\citenamefont{Loth et~al.}(2012)\citenamefont{Loth, Baumann, Lutz,
			Eigler, and Heinrich}}]{Loth13012012}
	\bibinfo{author}{\bibfnamefont{S.}~\bibnamefont{Loth}},
	\bibinfo{author}{\bibfnamefont{S.}~\bibnamefont{Baumann}},
	\bibinfo{author}{\bibfnamefont{C.~P.} \bibnamefont{Lutz}},
	\bibinfo{author}{\bibfnamefont{D.~M.} \bibnamefont{Eigler}},
	\bibnamefont{and} \bibinfo{author}{\bibfnamefont{A.~J.}
		\bibnamefont{Heinrich}}, \bibinfo{journal}{Science}
	\textbf{\bibinfo{volume}{335}}, \bibinfo{pages}{196} (\bibinfo{year}{2012}).
	
	\bibitem[{\citenamefont{Khajetoorians et~al.}(2013)\citenamefont{Khajetoorians,
			Baxevanis, H{\"u}bner, Schlenk, Krause, Wehling, Lounis, Lichtenstein,
			Pfannkuche, Wiebe et~al.}}]{Khajetoorians04012013}
	\bibinfo{author}{\bibfnamefont{A.~A.} \bibnamefont{Khajetoorians}},
	\bibinfo{author}{\bibfnamefont{B.}~\bibnamefont{Baxevanis}},
	\bibinfo{author}{\bibfnamefont{C.}~\bibnamefont{H{\"u}bner}},
	\bibinfo{author}{\bibfnamefont{T.}~\bibnamefont{Schlenk}},
	\bibinfo{author}{\bibfnamefont{S.}~\bibnamefont{Krause}},
	\bibinfo{author}{\bibfnamefont{T.~O.} \bibnamefont{Wehling}},
	\bibinfo{author}{\bibfnamefont{S.}~\bibnamefont{Lounis}},
	\bibinfo{author}{\bibfnamefont{A.}~\bibnamefont{Lichtenstein}},
	\bibinfo{author}{\bibfnamefont{D.}~\bibnamefont{Pfannkuche}},
	\bibinfo{author}{\bibfnamefont{J.}~\bibnamefont{Wiebe}},
	\bibnamefont{et~al.}, \bibinfo{journal}{Science}
	\textbf{\bibinfo{volume}{339}}, \bibinfo{pages}{55} (\bibinfo{year}{2013}).
	
	\bibitem[{\citenamefont{Wende et~al.}(2007)\citenamefont{Wende, Bernien, Luo,
			Sorg, Ponpandian, Kurde, Miguel, Piantek, Xu, Eckhold et~al.}}]{Wende:2007aa}
	\bibinfo{author}{\bibfnamefont{H.}~\bibnamefont{Wende}},
	\bibinfo{author}{\bibfnamefont{M.}~\bibnamefont{Bernien}},
	\bibinfo{author}{\bibfnamefont{J.}~\bibnamefont{Luo}},
	\bibinfo{author}{\bibfnamefont{C.}~\bibnamefont{Sorg}},
	\bibinfo{author}{\bibfnamefont{N.}~\bibnamefont{Ponpandian}},
	\bibinfo{author}{\bibfnamefont{J.}~\bibnamefont{Kurde}},
	\bibinfo{author}{\bibfnamefont{J.}~\bibnamefont{Miguel}},
	\bibinfo{author}{\bibfnamefont{M.}~\bibnamefont{Piantek}},
	\bibinfo{author}{\bibfnamefont{X.}~\bibnamefont{Xu}},
	\bibinfo{author}{\bibfnamefont{P.}~\bibnamefont{Eckhold}},
	\bibnamefont{et~al.}, \bibinfo{journal}{Nat Mater}
	\textbf{\bibinfo{volume}{6}}, \bibinfo{pages}{516} (\bibinfo{year}{2007}).
	
	\bibitem[{\citenamefont{Fern\'andez-Torrente
			et~al.}(2008)\citenamefont{Fern\'andez-Torrente, Franke, and
			Pascual}}]{PhysRevLett.101.217203}
	\bibinfo{author}{\bibfnamefont{I.}~\bibnamefont{Fern\'andez-Torrente}},
	\bibinfo{author}{\bibfnamefont{K.~J.} \bibnamefont{Franke}},
	\bibnamefont{and} \bibinfo{author}{\bibfnamefont{J.~I.}
		\bibnamefont{Pascual}}, \bibinfo{journal}{Phys. Rev. Lett.}
	\textbf{\bibinfo{volume}{101}}, \bibinfo{pages}{217203}
	(\bibinfo{year}{2008}).
	
	\bibitem[{\citenamefont{Chiesa et~al.}(2013)\citenamefont{Chiesa, Carretta,
			Santini, Amoretti, and Pavarini}}]{PhysRevLett.110.157204}
	\bibinfo{author}{\bibfnamefont{A.}~\bibnamefont{Chiesa}},
	\bibinfo{author}{\bibfnamefont{S.}~\bibnamefont{Carretta}},
	\bibinfo{author}{\bibfnamefont{P.}~\bibnamefont{Santini}},
	\bibinfo{author}{\bibfnamefont{G.}~\bibnamefont{Amoretti}}, \bibnamefont{and}
	\bibinfo{author}{\bibfnamefont{E.}~\bibnamefont{Pavarini}},
	\bibinfo{journal}{Phys. Rev. Lett.} \textbf{\bibinfo{volume}{110}},
	\bibinfo{pages}{157204} (\bibinfo{year}{2013}).
	
	\bibitem[{\citenamefont{Raman et~al.}(2013)\citenamefont{Raman, Kamerbeek,
			Mukherjee, Atodiresei, Sen, Lazic, Caciuc, Michel, Stalke, Mandal
			et~al.}}]{Raman:2013aa}
	\bibinfo{author}{\bibfnamefont{K.~V.} \bibnamefont{Raman}},
	\bibinfo{author}{\bibfnamefont{A.~M.} \bibnamefont{Kamerbeek}},
	\bibinfo{author}{\bibfnamefont{A.}~\bibnamefont{Mukherjee}},
	\bibinfo{author}{\bibfnamefont{N.}~\bibnamefont{Atodiresei}},
	\bibinfo{author}{\bibfnamefont{T.~K.} \bibnamefont{Sen}},
	\bibinfo{author}{\bibfnamefont{P.}~\bibnamefont{Lazic}},
	\bibinfo{author}{\bibfnamefont{V.}~\bibnamefont{Caciuc}},
	\bibinfo{author}{\bibfnamefont{R.}~\bibnamefont{Michel}},
	\bibinfo{author}{\bibfnamefont{D.}~\bibnamefont{Stalke}},
	\bibinfo{author}{\bibfnamefont{S.~K.} \bibnamefont{Mandal}},
	\bibnamefont{et~al.}, \bibinfo{journal}{Nature}
	\textbf{\bibinfo{volume}{493}}, \bibinfo{pages}{509} (\bibinfo{year}{2013}).
	
	\bibitem[{\citenamefont{Fahrendorf et~al.}(2013)\citenamefont{Fahrendorf,
			Atodiresei, Besson, Caciuc, Matthes, Bl{\"u}gel, K{\"o}gerler, B{\"u}rgler,
			and Schneider}}]{Fahrendorf:2013aa}
	\bibinfo{author}{\bibfnamefont{S.}~\bibnamefont{Fahrendorf}},
	\bibinfo{author}{\bibfnamefont{N.}~\bibnamefont{Atodiresei}},
	\bibinfo{author}{\bibfnamefont{C.}~\bibnamefont{Besson}},
	\bibinfo{author}{\bibfnamefont{V.}~\bibnamefont{Caciuc}},
	\bibinfo{author}{\bibfnamefont{F.}~\bibnamefont{Matthes}},
	\bibinfo{author}{\bibfnamefont{S.}~\bibnamefont{Bl{\"u}gel}},
	\bibinfo{author}{\bibfnamefont{P.}~\bibnamefont{K{\"o}gerler}},
	\bibinfo{author}{\bibfnamefont{D.~E.} \bibnamefont{B{\"u}rgler}},
	\bibnamefont{and} \bibinfo{author}{\bibfnamefont{C.~M.}
		\bibnamefont{Schneider}}, \bibinfo{journal}{Nat Commun}
	\textbf{\bibinfo{volume}{4}} (\bibinfo{year}{2013}).
	
	\bibitem[{\citenamefont{Carretta et~al.}(2004)\citenamefont{Carretta, Santini,
			Amoretti, Guidi, Caciuffo, Candini, Cornia, Gatteschi, Plazanet, and
			Stride}}]{PhysRevB.70.214403}
	\bibinfo{author}{\bibfnamefont{S.}~\bibnamefont{Carretta}},
	\bibinfo{author}{\bibfnamefont{P.}~\bibnamefont{Santini}},
	\bibinfo{author}{\bibfnamefont{G.}~\bibnamefont{Amoretti}},
	\bibinfo{author}{\bibfnamefont{T.}~\bibnamefont{Guidi}},
	\bibinfo{author}{\bibfnamefont{R.}~\bibnamefont{Caciuffo}},
	\bibinfo{author}{\bibfnamefont{A.}~\bibnamefont{Candini}},
	\bibinfo{author}{\bibfnamefont{A.}~\bibnamefont{Cornia}},
	\bibinfo{author}{\bibfnamefont{D.}~\bibnamefont{Gatteschi}},
	\bibinfo{author}{\bibfnamefont{M.}~\bibnamefont{Plazanet}}, \bibnamefont{and}
	\bibinfo{author}{\bibfnamefont{J.~A.} \bibnamefont{Stride}},
	\bibinfo{journal}{Phys. Rev. B} \textbf{\bibinfo{volume}{70}},
	\bibinfo{pages}{214403} (\bibinfo{year}{2004}).
	
	\bibitem[{\citenamefont{Cornia et~al.}(2004)\citenamefont{Cornia, Fabretti,
			Garrisi, Mortal{\`o}, Bonacchi, Gatteschi, Sessoli, Sorace, Wernsdorfer, and
			Barra}}]{ANIE:ANIE200352989}
	\bibinfo{author}{\bibfnamefont{A.}~\bibnamefont{Cornia}},
	\bibinfo{author}{\bibfnamefont{A.~C.} \bibnamefont{Fabretti}},
	\bibinfo{author}{\bibfnamefont{P.}~\bibnamefont{Garrisi}},
	\bibinfo{author}{\bibfnamefont{C.}~\bibnamefont{Mortal{\`o}}},
	\bibinfo{author}{\bibfnamefont{D.}~\bibnamefont{Bonacchi}},
	\bibinfo{author}{\bibfnamefont{D.}~\bibnamefont{Gatteschi}},
	\bibinfo{author}{\bibfnamefont{R.}~\bibnamefont{Sessoli}},
	\bibinfo{author}{\bibfnamefont{L.}~\bibnamefont{Sorace}},
	\bibinfo{author}{\bibfnamefont{W.}~\bibnamefont{Wernsdorfer}},
	\bibnamefont{and} \bibinfo{author}{\bibfnamefont{A.-L.} \bibnamefont{Barra}},
	\bibinfo{journal}{Angewandte Chemie International Edition}
	\textbf{\bibinfo{volume}{43}}, \bibinfo{pages}{1136} (\bibinfo{year}{2004}),
	ISSN \bibinfo{issn}{1521-3773}.
	
	\bibitem[{\citenamefont{van Slageren et~al.}(2002)\citenamefont{van Slageren,
			Sessoli, Gatteschi, Smith, Helliwell, Winpenny, Cornia, Barra, Jansen,
			Rentschler et~al.}}]{CHEM:CHEM277}
	\bibinfo{author}{\bibfnamefont{J.}~\bibnamefont{van Slageren}},
	\bibinfo{author}{\bibfnamefont{R.}~\bibnamefont{Sessoli}},
	\bibinfo{author}{\bibfnamefont{D.}~\bibnamefont{Gatteschi}},
	\bibinfo{author}{\bibfnamefont{A.~A.} \bibnamefont{Smith}},
	\bibinfo{author}{\bibfnamefont{M.}~\bibnamefont{Helliwell}},
	\bibinfo{author}{\bibfnamefont{R.~E.~P.} \bibnamefont{Winpenny}},
	\bibinfo{author}{\bibfnamefont{A.}~\bibnamefont{Cornia}},
	\bibinfo{author}{\bibfnamefont{A.-L.} \bibnamefont{Barra}},
	\bibinfo{author}{\bibfnamefont{A.~G.~M.} \bibnamefont{Jansen}},
	\bibinfo{author}{\bibfnamefont{E.}~\bibnamefont{Rentschler}},
	\bibnamefont{et~al.}, \bibinfo{journal}{Chemistry -- A European Journal}
	\textbf{\bibinfo{volume}{8}}, \bibinfo{pages}{277} (\bibinfo{year}{2002}),
	ISSN \bibinfo{issn}{1521-3765}.
	
	\bibitem[{\citenamefont{Carretta et~al.}(2003)\citenamefont{Carretta, van
			Slageren, Guidi, Liviotti, Mondelli, Rovai, Cornia, Dearden, Carsughi,
			Affronte et~al.}}]{PhysRevB.67.094405}
	\bibinfo{author}{\bibfnamefont{S.}~\bibnamefont{Carretta}},
	\bibinfo{author}{\bibfnamefont{J.}~\bibnamefont{van Slageren}},
	\bibinfo{author}{\bibfnamefont{T.}~\bibnamefont{Guidi}},
	\bibinfo{author}{\bibfnamefont{E.}~\bibnamefont{Liviotti}},
	\bibinfo{author}{\bibfnamefont{C.}~\bibnamefont{Mondelli}},
	\bibinfo{author}{\bibfnamefont{D.}~\bibnamefont{Rovai}},
	\bibinfo{author}{\bibfnamefont{A.}~\bibnamefont{Cornia}},
	\bibinfo{author}{\bibfnamefont{A.~L.} \bibnamefont{Dearden}},
	\bibinfo{author}{\bibfnamefont{F.}~\bibnamefont{Carsughi}},
	\bibinfo{author}{\bibfnamefont{M.}~\bibnamefont{Affronte}},
	\bibnamefont{et~al.}, \bibinfo{journal}{Phys. Rev. B}
	\textbf{\bibinfo{volume}{67}}, \bibinfo{pages}{094405}
	(\bibinfo{year}{2003}).
	
	\bibitem[{\citenamefont{Troiani et~al.}(2005)\citenamefont{Troiani, Ghirri,
			Affronte, Carretta, Santini, Amoretti, Piligkos, Timco, and
			Winpenny}}]{PhysRevLett.94.207208}
	\bibinfo{author}{\bibfnamefont{F.}~\bibnamefont{Troiani}},
	\bibinfo{author}{\bibfnamefont{A.}~\bibnamefont{Ghirri}},
	\bibinfo{author}{\bibfnamefont{M.}~\bibnamefont{Affronte}},
	\bibinfo{author}{\bibfnamefont{S.}~\bibnamefont{Carretta}},
	\bibinfo{author}{\bibfnamefont{P.}~\bibnamefont{Santini}},
	\bibinfo{author}{\bibfnamefont{G.}~\bibnamefont{Amoretti}},
	\bibinfo{author}{\bibfnamefont{S.}~\bibnamefont{Piligkos}},
	\bibinfo{author}{\bibfnamefont{G.}~\bibnamefont{Timco}}, \bibnamefont{and}
	\bibinfo{author}{\bibfnamefont{R.~E.~P.} \bibnamefont{Winpenny}},
	\bibinfo{journal}{Phys. Rev. Lett.} \textbf{\bibinfo{volume}{94}},
	\bibinfo{pages}{207208} (\bibinfo{year}{2005}).
	
	\bibitem[{\citenamefont{Carretta et~al.}(2007)\citenamefont{Carretta, Santini,
			Amoretti, Guidi, Copley, Qiu, Caciuffo, Timco, and
			Winpenny}}]{PhysRevLett.98.167401}
	\bibinfo{author}{\bibfnamefont{S.}~\bibnamefont{Carretta}},
	\bibinfo{author}{\bibfnamefont{P.}~\bibnamefont{Santini}},
	\bibinfo{author}{\bibfnamefont{G.}~\bibnamefont{Amoretti}},
	\bibinfo{author}{\bibfnamefont{T.}~\bibnamefont{Guidi}},
	\bibinfo{author}{\bibfnamefont{J.~R.~D.} \bibnamefont{Copley}},
	\bibinfo{author}{\bibfnamefont{Y.}~\bibnamefont{Qiu}},
	\bibinfo{author}{\bibfnamefont{R.}~\bibnamefont{Caciuffo}},
	\bibinfo{author}{\bibfnamefont{G.}~\bibnamefont{Timco}}, \bibnamefont{and}
	\bibinfo{author}{\bibfnamefont{R.~E.~P.} \bibnamefont{Winpenny}},
	\bibinfo{journal}{Phys. Rev. Lett.} \textbf{\bibinfo{volume}{98}},
	\bibinfo{pages}{167401} (\bibinfo{year}{2007}).
	
	\bibitem[{\citenamefont{Wedge et~al.}(2012)\citenamefont{Wedge, Timco,
			Spielberg, George, Tuna, Rigby, McInnes, Winpenny, Blundell, and
			Ardavan}}]{PhysRevLett.108.107204}
	\bibinfo{author}{\bibfnamefont{C.~J.} \bibnamefont{Wedge}},
	\bibinfo{author}{\bibfnamefont{G.~A.} \bibnamefont{Timco}},
	\bibinfo{author}{\bibfnamefont{E.~T.} \bibnamefont{Spielberg}},
	\bibinfo{author}{\bibfnamefont{R.~E.} \bibnamefont{George}},
	\bibinfo{author}{\bibfnamefont{F.}~\bibnamefont{Tuna}},
	\bibinfo{author}{\bibfnamefont{S.}~\bibnamefont{Rigby}},
	\bibinfo{author}{\bibfnamefont{E.~J.~L.} \bibnamefont{McInnes}},
	\bibinfo{author}{\bibfnamefont{R.~E.~P.} \bibnamefont{Winpenny}},
	\bibinfo{author}{\bibfnamefont{S.~J.} \bibnamefont{Blundell}},
	\bibnamefont{and} \bibinfo{author}{\bibfnamefont{A.}~\bibnamefont{Ardavan}},
	\bibinfo{journal}{Phys. Rev. Lett.} \textbf{\bibinfo{volume}{108}},
	\bibinfo{pages}{107204} (\bibinfo{year}{2012}).
	
	\bibitem[{\citenamefont{Candini et~al.}(2010)\citenamefont{Candini, Lorusso,
			Troiani, Ghirri, Carretta, Santini, Amoretti, Muryn, Tuna, Timco
			et~al.}}]{PhysRevLett.104.037203}
	\bibinfo{author}{\bibfnamefont{A.}~\bibnamefont{Candini}},
	\bibinfo{author}{\bibfnamefont{G.}~\bibnamefont{Lorusso}},
	\bibinfo{author}{\bibfnamefont{F.}~\bibnamefont{Troiani}},
	\bibinfo{author}{\bibfnamefont{A.}~\bibnamefont{Ghirri}},
	\bibinfo{author}{\bibfnamefont{S.}~\bibnamefont{Carretta}},
	\bibinfo{author}{\bibfnamefont{P.}~\bibnamefont{Santini}},
	\bibinfo{author}{\bibfnamefont{G.}~\bibnamefont{Amoretti}},
	\bibinfo{author}{\bibfnamefont{C.}~\bibnamefont{Muryn}},
	\bibinfo{author}{\bibfnamefont{F.}~\bibnamefont{Tuna}},
	\bibinfo{author}{\bibfnamefont{G.}~\bibnamefont{Timco}},
	\bibnamefont{et~al.}, \bibinfo{journal}{Phys. Rev. Lett.}
	\textbf{\bibinfo{volume}{104}}, \bibinfo{pages}{037203}
	(\bibinfo{year}{2010}).
	
	\bibitem[{\citenamefont{Chappert et~al.}(2007)\citenamefont{Chappert, Fert, and
			Van~Dau}}]{Chappert:2007aa}
	\bibinfo{author}{\bibfnamefont{C.}~\bibnamefont{Chappert}},
	\bibinfo{author}{\bibfnamefont{A.}~\bibnamefont{Fert}}, \bibnamefont{and}
	\bibinfo{author}{\bibfnamefont{F.~N.} \bibnamefont{Van~Dau}},
	\bibinfo{journal}{Nat Mater} \textbf{\bibinfo{volume}{6}},
	\bibinfo{pages}{813} (\bibinfo{year}{2007}).
	
	\bibitem[{\citenamefont{Locatelli et~al.}(2014)\citenamefont{Locatelli, Cros,
			and Grollier}}]{Locatelli:2014aa}
	\bibinfo{author}{\bibfnamefont{N.}~\bibnamefont{Locatelli}},
	\bibinfo{author}{\bibfnamefont{V.}~\bibnamefont{Cros}}, \bibnamefont{and}
	\bibinfo{author}{\bibfnamefont{J.}~\bibnamefont{Grollier}},
	\bibinfo{journal}{Nat Mater} \textbf{\bibinfo{volume}{13}},
	\bibinfo{pages}{11} (\bibinfo{year}{2014}).
	
	\bibitem[{\citenamefont{Bogani and Wernsdorfer}(2008)}]{Bogani:2008aa}
	\bibinfo{author}{\bibfnamefont{L.}~\bibnamefont{Bogani}} \bibnamefont{and}
	\bibinfo{author}{\bibfnamefont{W.}~\bibnamefont{Wernsdorfer}},
	\bibinfo{journal}{Nat Mater} \textbf{\bibinfo{volume}{7}},
	\bibinfo{pages}{179} (\bibinfo{year}{2008}).
	
	\bibitem[{\citenamefont{Leuenberger and Loss}(2001)}]{Leuenberger:2001aa}
	\bibinfo{author}{\bibfnamefont{M.~N.} \bibnamefont{Leuenberger}}
	\bibnamefont{and} \bibinfo{author}{\bibfnamefont{D.}~\bibnamefont{Loss}},
	\bibinfo{journal}{Nature} \textbf{\bibinfo{volume}{410}},
	\bibinfo{pages}{789} (\bibinfo{year}{2001}).
	
	\bibitem[{\citenamefont{Antropov et~al.}(1995)\citenamefont{Antropov,
			Katsnelson, van Schilfgaarde, and Harmon}}]{PhysRevLett.75.729}
	\bibinfo{author}{\bibfnamefont{V.~P.} \bibnamefont{Antropov}},
	\bibinfo{author}{\bibfnamefont{M.~I.} \bibnamefont{Katsnelson}},
	\bibinfo{author}{\bibfnamefont{M.}~\bibnamefont{van Schilfgaarde}},
	\bibnamefont{and} \bibinfo{author}{\bibfnamefont{B.~N.}
		\bibnamefont{Harmon}}, \bibinfo{journal}{Phys. Rev. Lett.}
	\textbf{\bibinfo{volume}{75}}, \bibinfo{pages}{729} (\bibinfo{year}{1995}).
	
	\bibitem[{\citenamefont{Mannini et~al.}(2009)\citenamefont{Mannini, Pineider,
			Sainctavit, Danieli, Otero, Sciancalepore, Talarico, Arrio, Cornia, Gatteschi
			et~al.}}]{Mannini:2009aa}
	\bibinfo{author}{\bibfnamefont{M.}~\bibnamefont{Mannini}},
	\bibinfo{author}{\bibfnamefont{F.}~\bibnamefont{Pineider}},
	\bibinfo{author}{\bibfnamefont{P.}~\bibnamefont{Sainctavit}},
	\bibinfo{author}{\bibfnamefont{C.}~\bibnamefont{Danieli}},
	\bibinfo{author}{\bibfnamefont{E.}~\bibnamefont{Otero}},
	\bibinfo{author}{\bibfnamefont{C.}~\bibnamefont{Sciancalepore}},
	\bibinfo{author}{\bibfnamefont{A.~M.} \bibnamefont{Talarico}},
	\bibinfo{author}{\bibfnamefont{M.-A.} \bibnamefont{Arrio}},
	\bibinfo{author}{\bibfnamefont{A.}~\bibnamefont{Cornia}},
	\bibinfo{author}{\bibfnamefont{D.}~\bibnamefont{Gatteschi}},
	\bibnamefont{et~al.}, \bibinfo{journal}{Nat Mater}
	\textbf{\bibinfo{volume}{8}}, \bibinfo{pages}{194} (\bibinfo{year}{2009}).
	
	\bibitem[{\citenamefont{Timco et~al.}(2009)\citenamefont{Timco, Carretta,
			Troiani, Tuna, Pritchard, Muryn, McInnes, Ghirri, Candini, Santini
			et~al.}}]{Timco:2009aa}
	\bibinfo{author}{\bibfnamefont{G.~A.} \bibnamefont{Timco}},
	\bibinfo{author}{\bibfnamefont{S.}~\bibnamefont{Carretta}},
	\bibinfo{author}{\bibfnamefont{F.}~\bibnamefont{Troiani}},
	\bibinfo{author}{\bibfnamefont{F.}~\bibnamefont{Tuna}},
	\bibinfo{author}{\bibfnamefont{R.~J.} \bibnamefont{Pritchard}},
	\bibinfo{author}{\bibfnamefont{C.~A.} \bibnamefont{Muryn}},
	\bibinfo{author}{\bibfnamefont{E.~J.~L.} \bibnamefont{McInnes}},
	\bibinfo{author}{\bibfnamefont{A.}~\bibnamefont{Ghirri}},
	\bibinfo{author}{\bibfnamefont{A.}~\bibnamefont{Candini}},
	\bibinfo{author}{\bibfnamefont{P.}~\bibnamefont{Santini}},
	\bibnamefont{et~al.}, \bibinfo{journal}{Nat Nano}
	\textbf{\bibinfo{volume}{4}}, \bibinfo{pages}{173} (\bibinfo{year}{2009}).
	
	\bibitem[{\citenamefont{Mannini et~al.}(2010)\citenamefont{Mannini, Pineider,
			Danieli, Totti, Sorace, Sainctavit, Arrio, Otero, Joly, Cezar
			et~al.}}]{Mannini:2010aa}
	\bibinfo{author}{\bibfnamefont{M.}~\bibnamefont{Mannini}},
	\bibinfo{author}{\bibfnamefont{F.}~\bibnamefont{Pineider}},
	\bibinfo{author}{\bibfnamefont{C.}~\bibnamefont{Danieli}},
	\bibinfo{author}{\bibfnamefont{F.}~\bibnamefont{Totti}},
	\bibinfo{author}{\bibfnamefont{L.}~\bibnamefont{Sorace}},
	\bibinfo{author}{\bibfnamefont{P.}~\bibnamefont{Sainctavit}},
	\bibinfo{author}{\bibfnamefont{M.~A.} \bibnamefont{Arrio}},
	\bibinfo{author}{\bibfnamefont{E.}~\bibnamefont{Otero}},
	\bibinfo{author}{\bibfnamefont{L.}~\bibnamefont{Joly}},
	\bibinfo{author}{\bibfnamefont{J.~C.} \bibnamefont{Cezar}},
	\bibnamefont{et~al.}, \bibinfo{journal}{Nature}
	\textbf{\bibinfo{volume}{468}}, \bibinfo{pages}{417} (\bibinfo{year}{2010}).
	
	\bibitem[{\citenamefont{Carretta et~al.}(2006)\citenamefont{Carretta, Santini,
			Amoretti, Affronte, Candini, Ghirri, Tidmarsh, Laye, Shaw, and
			McInnes}}]{PhysRevLett.97.207201}
	\bibinfo{author}{\bibfnamefont{S.}~\bibnamefont{Carretta}},
	\bibinfo{author}{\bibfnamefont{P.}~\bibnamefont{Santini}},
	\bibinfo{author}{\bibfnamefont{G.}~\bibnamefont{Amoretti}},
	\bibinfo{author}{\bibfnamefont{M.}~\bibnamefont{Affronte}},
	\bibinfo{author}{\bibfnamefont{A.}~\bibnamefont{Candini}},
	\bibinfo{author}{\bibfnamefont{A.}~\bibnamefont{Ghirri}},
	\bibinfo{author}{\bibfnamefont{I.~S.} \bibnamefont{Tidmarsh}},
	\bibinfo{author}{\bibfnamefont{R.~H.} \bibnamefont{Laye}},
	\bibinfo{author}{\bibfnamefont{R.}~\bibnamefont{Shaw}}, \bibnamefont{and}
	\bibinfo{author}{\bibfnamefont{E.~J.~L.} \bibnamefont{McInnes}},
	\bibinfo{journal}{Phys. Rev. Lett.} \textbf{\bibinfo{volume}{97}},
	\bibinfo{pages}{207201} (\bibinfo{year}{2006}).
	
	\bibitem[{\citenamefont{Bhattacharjee et~al.}(2012)\citenamefont{Bhattacharjee,
			Nordstr\"om, and Fransson}}]{PhysRevLett.108.057204}
	\bibinfo{author}{\bibfnamefont{S.}~\bibnamefont{Bhattacharjee}},
	\bibinfo{author}{\bibfnamefont{L.}~\bibnamefont{Nordstr\"om}},
	\bibnamefont{and} \bibinfo{author}{\bibfnamefont{J.}~\bibnamefont{Fransson}},
	\bibinfo{journal}{Phys. Rev. Lett.} \textbf{\bibinfo{volume}{108}},
	\bibinfo{pages}{057204} (\bibinfo{year}{2012}).
	
	\bibitem[{\citenamefont{Zhang and Zhang}(2009)}]{PhysRevLett.102.086601}
	\bibinfo{author}{\bibfnamefont{S.}~\bibnamefont{Zhang}} \bibnamefont{and}
	\bibinfo{author}{\bibfnamefont{S.~S.-L.} \bibnamefont{Zhang}},
	\bibinfo{journal}{Phys. Rev. Lett.} \textbf{\bibinfo{volume}{102}},
	\bibinfo{pages}{086601} (\bibinfo{year}{2009}).
	
	\bibitem[{\citenamefont{Szilva et~al.}(2013)\citenamefont{Szilva, Costa,
			Bergman, Szunyogh, Nordstr\"om, and Eriksson}}]{PhysRevLett.111.127204}
	\bibinfo{author}{\bibfnamefont{A.}~\bibnamefont{Szilva}},
	\bibinfo{author}{\bibfnamefont{M.}~\bibnamefont{Costa}},
	\bibinfo{author}{\bibfnamefont{A.}~\bibnamefont{Bergman}},
	\bibinfo{author}{\bibfnamefont{L.}~\bibnamefont{Szunyogh}},
	\bibinfo{author}{\bibfnamefont{L.}~\bibnamefont{Nordstr\"om}},
	\bibnamefont{and} \bibinfo{author}{\bibfnamefont{O.}~\bibnamefont{Eriksson}},
	\bibinfo{journal}{Phys. Rev. Lett.} \textbf{\bibinfo{volume}{111}},
	\bibinfo{pages}{127204} (\bibinfo{year}{2013}).
	
	\bibitem[{\citenamefont{Fransson}(2008{\natexlab{a}})}]{PhysRevB.77.205316}
	\bibinfo{author}{\bibfnamefont{J.}~\bibnamefont{Fransson}},
	\bibinfo{journal}{Phys. Rev. B} \textbf{\bibinfo{volume}{77}},
	\bibinfo{pages}{205316} (\bibinfo{year}{2008}{\natexlab{a}}).
	
	\bibitem[{\citenamefont{Filipovi\ifmmode~\acute{c}\else \'{c}\fi{}
			et~al.}(2013)\citenamefont{Filipovi\ifmmode~\acute{c}\else \'{c}\fi{},
			Holmqvist, Haupt, and Belzig}}]{PhysRevB.87.045426}
	\bibinfo{author}{\bibfnamefont{M.}~\bibnamefont{Filipovi\ifmmode~\acute{c}\else
			\'{c}\fi{}}}, \bibinfo{author}{\bibfnamefont{C.}~\bibnamefont{Holmqvist}},
	\bibinfo{author}{\bibfnamefont{F.}~\bibnamefont{Haupt}}, \bibnamefont{and}
	\bibinfo{author}{\bibfnamefont{W.}~\bibnamefont{Belzig}},
	\bibinfo{journal}{Phys. Rev. B} \textbf{\bibinfo{volume}{87}},
	\bibinfo{pages}{045426} (\bibinfo{year}{2013}).
	
	\bibitem[{\citenamefont{Fransson et~al.}(2014)\citenamefont{Fransson, Ren, and
			Zhu}}]{PhysRevLett.113.257201}
	\bibinfo{author}{\bibfnamefont{J.}~\bibnamefont{Fransson}},
	\bibinfo{author}{\bibfnamefont{J.}~\bibnamefont{Ren}}, \bibnamefont{and}
	\bibinfo{author}{\bibfnamefont{J.-X.} \bibnamefont{Zhu}},
	\bibinfo{journal}{Phys. Rev. Lett.} \textbf{\bibinfo{volume}{113}},
	\bibinfo{pages}{257201} (\bibinfo{year}{2014}).
	
	\bibitem[{\citenamefont{Misiorny et~al.}(2013)\citenamefont{Misiorny, Hell, and
			Wegewijs}}]{Misiorny:2013aa}
	\bibinfo{author}{\bibfnamefont{M.}~\bibnamefont{Misiorny}},
	\bibinfo{author}{\bibfnamefont{M.}~\bibnamefont{Hell}}, \bibnamefont{and}
	\bibinfo{author}{\bibfnamefont{M.~R.} \bibnamefont{Wegewijs}},
	\bibinfo{journal}{Nat Phys} \textbf{\bibinfo{volume}{9}},
	\bibinfo{pages}{801} (\bibinfo{year}{2013}).
	
	\bibitem[{\citenamefont{Timm and Elste}(2006)}]{PhysRevB.73.235304}
	\bibinfo{author}{\bibfnamefont{C.}~\bibnamefont{Timm}} \bibnamefont{and}
	\bibinfo{author}{\bibfnamefont{F.}~\bibnamefont{Elste}},
	\bibinfo{journal}{Phys. Rev. B} \textbf{\bibinfo{volume}{73}},
	\bibinfo{pages}{235304} (\bibinfo{year}{2006}).
	
	\bibitem[{\citenamefont{Misiorny and Barna\ifmmode~\acute{s}\else
			\'{s}\fi{}}(2007)}]{PhysRevB.75.134425}
	\bibinfo{author}{\bibfnamefont{M.}~\bibnamefont{Misiorny}} \bibnamefont{and}
	\bibinfo{author}{\bibfnamefont{J.}~\bibnamefont{Barna\ifmmode~\acute{s}\else
			\'{s}\fi{}}}, \bibinfo{journal}{Phys. Rev. B} \textbf{\bibinfo{volume}{75}},
	\bibinfo{pages}{134425} (\bibinfo{year}{2007}).
	
	\bibitem[{\citenamefont{Moldoveanu et~al.}(2015)\citenamefont{Moldoveanu, Dinu,
			Tanatar, and Moca}}]{1367-2630-17-8-083020}
	\bibinfo{author}{\bibfnamefont{V.}~\bibnamefont{Moldoveanu}},
	\bibinfo{author}{\bibfnamefont{I.~V.} \bibnamefont{Dinu}},
	\bibinfo{author}{\bibfnamefont{B.}~\bibnamefont{Tanatar}}, \bibnamefont{and}
	\bibinfo{author}{\bibfnamefont{C.~P.} \bibnamefont{Moca}},
	\bibinfo{journal}{New Journal of Physics} \textbf{\bibinfo{volume}{17}},
	\bibinfo{pages}{083020} (\bibinfo{year}{2015}).
	
	\bibitem[{\citenamefont{H\"artle et~al.}(2011)\citenamefont{H\"artle, Butzin,
			Rubio-Pons, and Thoss}}]{PhysRevLett.107.046802}
	\bibinfo{author}{\bibfnamefont{R.}~\bibnamefont{H\"artle}},
	\bibinfo{author}{\bibfnamefont{M.}~\bibnamefont{Butzin}},
	\bibinfo{author}{\bibfnamefont{O.}~\bibnamefont{Rubio-Pons}},
	\bibnamefont{and} \bibinfo{author}{\bibfnamefont{M.}~\bibnamefont{Thoss}},
	\bibinfo{journal}{Phys. Rev. Lett.} \textbf{\bibinfo{volume}{107}},
	\bibinfo{pages}{046802} (\bibinfo{year}{2011}).
	
	\bibitem[{\citenamefont{Roura-Bas et~al.}(2013)\citenamefont{Roura-Bas, Tosi,
			and Aligia}}]{PhysRevB.87.195136}
	\bibinfo{author}{\bibfnamefont{P.}~\bibnamefont{Roura-Bas}},
	\bibinfo{author}{\bibfnamefont{L.}~\bibnamefont{Tosi}}, \bibnamefont{and}
	\bibinfo{author}{\bibfnamefont{A.~A.} \bibnamefont{Aligia}},
	\bibinfo{journal}{Phys. Rev. B} \textbf{\bibinfo{volume}{87}},
	\bibinfo{pages}{195136} (\bibinfo{year}{2013}).
	
	\bibitem[{\citenamefont{Linder and Robinson}(2015)}]{Linder:2015aa}
	\bibinfo{author}{\bibfnamefont{J.}~\bibnamefont{Linder}} \bibnamefont{and}
	\bibinfo{author}{\bibfnamefont{J.~W.~A.} \bibnamefont{Robinson}},
	\bibinfo{journal}{Nat Phys} \textbf{\bibinfo{volume}{11}},
	\bibinfo{pages}{307} (\bibinfo{year}{2015}).
	
	\bibitem[{\citenamefont{Stadler et~al.}(2013)\citenamefont{Stadler, Holmqvist,
			and Belzig}}]{PhysRevB.88.104512}
	\bibinfo{author}{\bibfnamefont{P.}~\bibnamefont{Stadler}},
	\bibinfo{author}{\bibfnamefont{C.}~\bibnamefont{Holmqvist}},
	\bibnamefont{and} \bibinfo{author}{\bibfnamefont{W.}~\bibnamefont{Belzig}},
	\bibinfo{journal}{Phys. Rev. B} \textbf{\bibinfo{volume}{88}},
	\bibinfo{pages}{104512} (\bibinfo{year}{2013}).
	
	\bibitem[{\citenamefont{Holmqvist et~al.}(2011)\citenamefont{Holmqvist, Teber,
			and Fogelstr\"om}}]{PhysRevB.83.104521}
	\bibinfo{author}{\bibfnamefont{C.}~\bibnamefont{Holmqvist}},
	\bibinfo{author}{\bibfnamefont{S.}~\bibnamefont{Teber}}, \bibnamefont{and}
	\bibinfo{author}{\bibfnamefont{M.}~\bibnamefont{Fogelstr\"om}},
	\bibinfo{journal}{Phys. Rev. B} \textbf{\bibinfo{volume}{83}},
	\bibinfo{pages}{104521} (\bibinfo{year}{2011}).
	
	\bibitem[{\citenamefont{Holmqvist et~al.}(2014)\citenamefont{Holmqvist,
			Fogelstr\"om, and Belzig}}]{PhysRevB.90.014516}
	\bibinfo{author}{\bibfnamefont{C.}~\bibnamefont{Holmqvist}},
	\bibinfo{author}{\bibfnamefont{M.}~\bibnamefont{Fogelstr\"om}},
	\bibnamefont{and} \bibinfo{author}{\bibfnamefont{W.}~\bibnamefont{Belzig}},
	\bibinfo{journal}{Phys. Rev. B} \textbf{\bibinfo{volume}{90}},
	\bibinfo{pages}{014516} (\bibinfo{year}{2014}).
	
	\bibitem[{\citenamefont{Fransson and Zhu}(2008)}]{franssonNJP2008}
	\bibinfo{author}{\bibfnamefont{J.}~\bibnamefont{Fransson}} \bibnamefont{and}
	\bibinfo{author}{\bibfnamefont{J.-X.} \bibnamefont{Zhu}},
	\bibinfo{journal}{New J. Phys.} \textbf{\bibinfo{volume}{10}},
	\bibinfo{pages}{013017} (\bibinfo{year}{2008}).
	
	\bibitem[{\citenamefont{Fransson}(2008{\natexlab{b}})}]{fransson2008}
	\bibinfo{author}{\bibfnamefont{J.}~\bibnamefont{Fransson}},
	\bibinfo{journal}{Nanotechnology} \textbf{\bibinfo{volume}{19}},
	\bibinfo{pages}{285714} (\bibinfo{year}{2008}{\natexlab{b}}).
	
	\bibitem[{\citenamefont{Bode et~al.}(2012)\citenamefont{Bode, Arrachea, Lozano,
			Nunner, and von Oppen}}]{PhysRevB.85.115440}
	\bibinfo{author}{\bibfnamefont{N.}~\bibnamefont{Bode}},
	\bibinfo{author}{\bibfnamefont{L.}~\bibnamefont{Arrachea}},
	\bibinfo{author}{\bibfnamefont{G.~S.} \bibnamefont{Lozano}},
	\bibinfo{author}{\bibfnamefont{T.~S.} \bibnamefont{Nunner}},
	\bibnamefont{and} \bibinfo{author}{\bibfnamefont{F.}~\bibnamefont{von
			Oppen}}, \bibinfo{journal}{Phys. Rev. B} \textbf{\bibinfo{volume}{85}},
	\bibinfo{pages}{115440} (\bibinfo{year}{2012}).
	
	\bibitem[{\citenamefont{Arrachea and von Oppen}(2015)}]{Arrachea:2015aa}
	\bibinfo{author}{\bibfnamefont{L.}~\bibnamefont{Arrachea}} \bibnamefont{and}
	\bibinfo{author}{\bibfnamefont{F.}~\bibnamefont{von Oppen}},
	\bibinfo{journal}{Physica E: Low-dimensional Systems and Nanostructures}
	\textbf{\bibinfo{volume}{74}}, \bibinfo{pages}{596} (\bibinfo{year}{2015}).
	
	\bibitem[{\citenamefont{Ralph and Stiles}(2008)}]{Ralph20081190}
	\bibinfo{author}{\bibfnamefont{D.}~\bibnamefont{Ralph}} \bibnamefont{and}
	\bibinfo{author}{\bibfnamefont{M.}~\bibnamefont{Stiles}},
	\bibinfo{journal}{Journal of Magnetism and Magnetic Materials}
	\textbf{\bibinfo{volume}{320}}, \bibinfo{pages}{1190 }
	(\bibinfo{year}{2008}), ISSN \bibinfo{issn}{0304-8853}.
	
	\bibitem[{\citenamefont{Wingreen et~al.}(1993)\citenamefont{Wingreen, Jauho,
			and Meir}}]{PhysRevB.48.8487}
	\bibinfo{author}{\bibfnamefont{N.~S.} \bibnamefont{Wingreen}},
	\bibinfo{author}{\bibfnamefont{A.-P.} \bibnamefont{Jauho}}, \bibnamefont{and}
	\bibinfo{author}{\bibfnamefont{Y.}~\bibnamefont{Meir}},
	\bibinfo{journal}{Phys. Rev. B} \textbf{\bibinfo{volume}{48}},
	\bibinfo{pages}{8487} (\bibinfo{year}{1993}).
	
	\bibitem[{\citenamefont{Jauho et~al.}(1994)\citenamefont{Jauho, Wingreen, and
			Meir}}]{PhysRevB.50.5528}
	\bibinfo{author}{\bibfnamefont{A.-P.} \bibnamefont{Jauho}},
	\bibinfo{author}{\bibfnamefont{N.~S.} \bibnamefont{Wingreen}},
	\bibnamefont{and} \bibinfo{author}{\bibfnamefont{Y.}~\bibnamefont{Meir}},
	\bibinfo{journal}{Phys. Rev. B} \textbf{\bibinfo{volume}{50}},
	\bibinfo{pages}{5528} (\bibinfo{year}{1994}).
	
	\bibitem[{\citenamefont{chao Chou et~al.}(1985)\citenamefont{chao Chou, bin Su,
			lin Hao, and Yu}}]{CHOU19851}
	\bibinfo{author}{\bibfnamefont{K.}~\bibnamefont{chao Chou}},
	\bibinfo{author}{\bibfnamefont{Z.}~\bibnamefont{bin Su}},
	\bibinfo{author}{\bibfnamefont{B.}~\bibnamefont{lin Hao}}, \bibnamefont{and}
	\bibinfo{author}{\bibfnamefont{L.}~\bibnamefont{Yu}},
	\bibinfo{journal}{Physics Reports} \textbf{\bibinfo{volume}{118}},
	\bibinfo{pages}{1 } (\bibinfo{year}{1985}), ISSN \bibinfo{issn}{0370-1573}.
	
\end{thebibliography}
\end{document}